\documentclass[prr,twocolumn,superscriptaddress,longbibliography]{revtex4-2}
\usepackage[english]{babel}
\usepackage{amsmath,amssymb,bm}
\usepackage{physics}
\usepackage{graphicx}
\usepackage{subfig}
\usepackage{tikz}
\usepackage{xcolor}
\usetikzlibrary{arrows.meta, decorations.pathmorphing}
\usepackage[colorlinks=true,linkcolor=blue,citecolor=blue,urlcolor=blue]{hyperref}
\usepackage{comment}

\begin{document}

\title{Exactly Solvable Phase Transition in a Cavity-Coupled One-Dimensional Ising Chain}

\author{Shuntaro Otake}
\email{shuntarootake@g.ecc.u-tokyo.ac.jp}
\affiliation{Department of Basic Science, The University of Tokyo, Komaba, Meguro-ku, Tokyo 153-8902, Japan
}
\affiliation{Department of Physics, College of Engineering Science, Yokohama National University, 79-5 Tokiwadai, Hodogaya-ku, Yokohama 240-8501, Japan}

\author{Motoaki Bamba}
\affiliation{Department of Physics, College of Engineering Science, Yokohama National University, 79-5 Tokiwadai, Hodogaya-ku, Yokohama 240-8501, Japan}
\affiliation{Department of Physics, Graduate School of Engineering Science, Yokohama National University, 79-5 Tokiwadai, Hodogaya-ku, Yokohama 240-8501, Japan}
\affiliation{Institute for Multidisciplinary Sciences, Yokohama National University, 79-5 Tokiwadai, Hodogaya-ku, Yokohama 240-8501, Japan}

\date{\today}

\begin{abstract}
Although one-dimensional classical spin chains do not exhibit phase transitions, we found that a phase transition does occur when they are coupled to a cavity photon mode. This provides the simplest exactly solvable examples demonstrating that finite-temperature superradiant phase transitions can emerge from long-range fully connected interactions mediated by photons and interactions within the material.
\end{abstract}

\maketitle

\subsection{Introduction}
A hundred years ago, Ising, in his doctoral dissertation, showed that a one-dimensional (1D) classical ferromagnetic spin chain does not exhibit a magnetic phase transition \cite{ising1925}. However, it is nowadays widely known that magnetic phase transitions occur in two or more dimensions, and since then the Ising model has been positioned as one of the simplest, crucial models for studying phase-transition phenomena in many-body systems. 

Furthermore, the strong interaction between matter and light, which is expected to be realized in cavity systems in recent years, is expected to lead to the discovery and understanding of new emergent phenomena. The ultrastrong light--matter coupling \cite{forn-diaz2019,kockum2019} has, over the past decade, moved beyond theoretical concepts and has been realized in a variety of experimental systems, including phonons \cite{Faust1966-rm,Zhang2021-rw,Barra-Burillo2021-hr}, inter-subband transitions in semiconductor quantum wells \cite{Gunter2009-yd}, superconducting circuits \cite{Niemczyk2010-jk,Yoshihara2017-fa}, organic molecules \cite{Schwartz2011-ir}, electron cyclotron motion (Landau-level transitions) \cite{Scalari2012-rw,Bayer2017-kb}, molecular vibrations \cite{George2016-kc}, surface plasmon polaritons \cite{Mueller2020-bo} and magnons \cite{Golovchanskiy2021-hi}. This progress has opened promising avenues for potential applications in both physics and chemistry \cite{forn-diaz2019,kockum2019,Nagarajan2021-rd}.

Of special interest is the realization of a superradiant phase transition (SRPT)~\cite{hepp1973}, namely, the spontaneous build-up of a coherent electromagnetic field under thermal equilibrium due to light–matter coupling. While lasers also display spontaneous coherence, they rely on population inversion and are therefore intrinsically nonequilibrium. The SRPT was first proposed in the Dicke model~\cite{hepp1973,dicke1954}, which treats non-interacting two-level atoms and thus neglects interactions within the medium; its realization in interacting settings remains under active study \cite{Rohn:2020:IML,Nakamoto2025,Langheld2025,Shapiro2025,Koziol2025,Cortese2017,Lee2004,Schellenberger2024,RomanRoche2025Bound, Mendonca2025}.
Notably, while recent studies on cavity-coupled 1D systems~\cite{Nakamoto2025, Langheld2025, Kirton2019} have focused on the behavior at absolute zero temperature, our work explores the exact properties at finite temperatures.

To this end, we present a minimal interacting route: an exactly solvable cavity-coupled \emph{classical} 1D Ising chain.
Consequently, an SRPT occurs, providing the simplest solvable interacting 1D example. This construction makes the mechanism explicit---\emph{at the Hamiltonian level}---showing how photon-mediated fully connected coupling, together with intrinsic interactions, produces finite-temperature superradiant order.

In this paper, we first discuss the general properties of cavity-coupled spin models. Subsequently, we investigate the 1D classical model. Because this model does not explicitly involve quantum fluctuations, it allows us to exactly reproduce the SRPT at arbitrary finite temperatures.

\begin{figure}[htbp]
  \centering
  \includegraphics[width=0.7\linewidth]{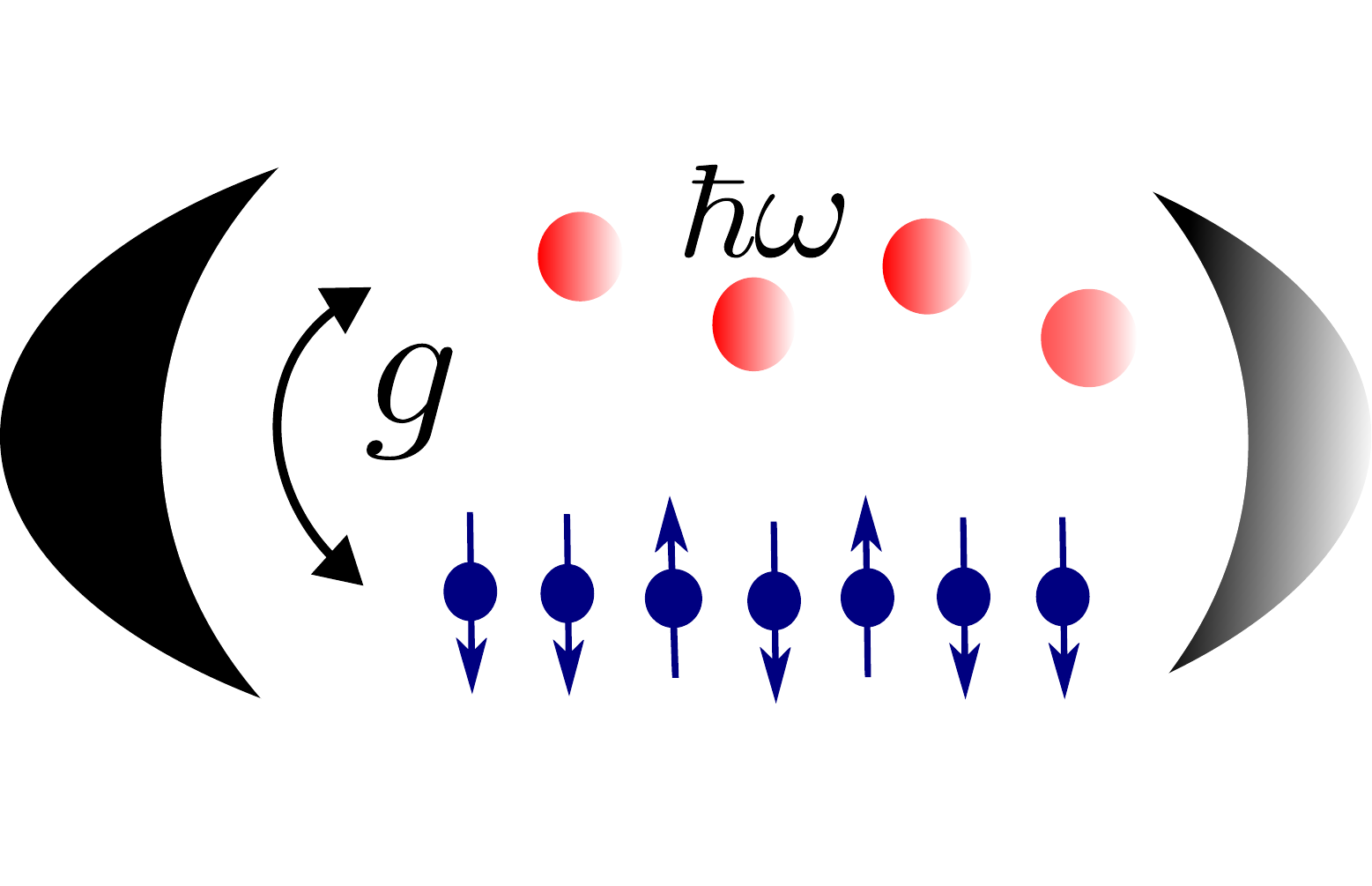}
  \caption{The spin system is coupled to cavity photons with a coupling constant $g$. The figure shows seven spins depicted in blue, coupled to photons illustrated in red.}
  \label{fig:diagram1}
\end{figure}

\subsection{Model}
We consider a typical lattice in which spins are located. Let the total number of spins be $N$. The spin at site $i\in \lbrace 1,2,\dots,N \rbrace$ is generally given by the set $(\sigma_i^x,\sigma_i^y,\sigma_i^z)$ where for each $i$, $\sigma_i^\mu\,(\mu=x,y,z)$ represents the Pauli matrices whose eigenvalues are $\pm 1$. 
We denote the Hamiltonian of the spin system as $H_\text{spin}$.
For example, a typical classical Ising model is given by
\begin{equation}
H_\text{Ising}= -\sum_{\langle i,j \rangle} J_{ij} \sigma_i^z \sigma_j^z. \label{Ising}
\end{equation}
where \(\langle i,j\rangle\) denotes nearest neighbors and \(J_{ij}\) is the (exchange) interaction.

We consider a spin system coupled to a single cavity mode,
\begin{equation}
H=H_{\rm spin}+\omega\,a^\dagger a+\frac{g}{\sqrt{N}}\,(a+a^\dagger)\sum_{i=1}^N\sigma_i^\mu,\quad \mu\in\{x,y,z\},
\label{model}
\end{equation}
with $\hbar=1$, cavity frequency $\omega$, bosonic operators $[a,a^\dagger]=1$ and coupling strength $g$ between light and spin ensemble. 
The spin component $\mu$ depends on the microscopic implementation; e.g., for the Dicke model~\cite{hepp1973,dicke1954} with noninteracting two-level atoms ($H_{\rm spin}=\omega_0\sum_i\sigma_i^z$) one has $\mu=x$:
$
H_{\rm Dicke}=\omega_0\sum_{i=1}^N\sigma_i^z+\omega a^\dagger a+\frac{g}{\sqrt{N}}(a+a^\dagger)\sum_{i=1}^N\sigma_i^x .
\label{Dicke}
$

We set $k_B=1$ and denote thermal averages by $\langle\cdot\rangle$. 
The order parameters used below are
\begin{equation}
m = \frac{1}{N}\sum_{i=1}^N \langle \sigma_i^\mu\rangle,
\qquad
n = \frac{1}{N}\langle a^\dagger a\rangle.
\end{equation}
Here, $m$ is the magnetization of the system in the $\mu$ direction, and $n$ is the photon number normalized by the number of spins ($N$), which corresponds to the square of the normalized cavity electromagnetic field in the thermodynamic limit ($N \to \infty$) as seen later in Eq.~\eqref{n}.

We diagonalize the light sector by a displacement of the cavity field.
\begin{equation}
b = a+\frac{g}{\omega\sqrt{N}}\sum_{i=1}^N\sigma_i^\mu,\qquad [b,b^\dagger]=1,
\end{equation}
which yields
\begin{align}
H
&=H_{\rm spin}+\omega\,b^\dagger b-\frac{g^2}{N\omega}\!\left(\sum_{i=1}^N\sigma_i^\mu\right)^{\!2} \nonumber\\
&=H_{\rm spin}-\frac{g^2}{N\omega}\sum_{i\neq j}\sigma_i^\mu\sigma_j^\mu+\omega\,b^\dagger b-\frac{g^2}{\omega}.
\label{H_d}
\end{align}
The generated residual term is constant due to $(\sigma_i^\mu)^2=1$.

 The photon mode associated with the annihilation operator $b$ is decoupled from the spins because the linear interaction can be removed by a unitary displacement transformation~\ref{sec:general_theory}. We obtain the following partition function:
\begin{equation}
    Z = \underbrace{\Tr_{\text{spin}} \left[ e^{-\beta H_{\text{eff}}} \right]}_{Z_{\text{spin}}} \times \underbrace{\Tr_{\text{phot}} \left[ e^{-\beta \omega a^{\dagger}a} \right]}_{Z_{\text{boson}}},
\end{equation}
where we defined an effective Hamiltonian $H_{\text{eff}} = H_{\text{spin}} - \frac{g^{2}}{N\omega}\left(\sum_{i=1}^{N}\sigma_{i}^{\mu}\right)^{2}$.

\begin{figure}[htbp] 
  \centering
  \includegraphics[width=0.83\linewidth]{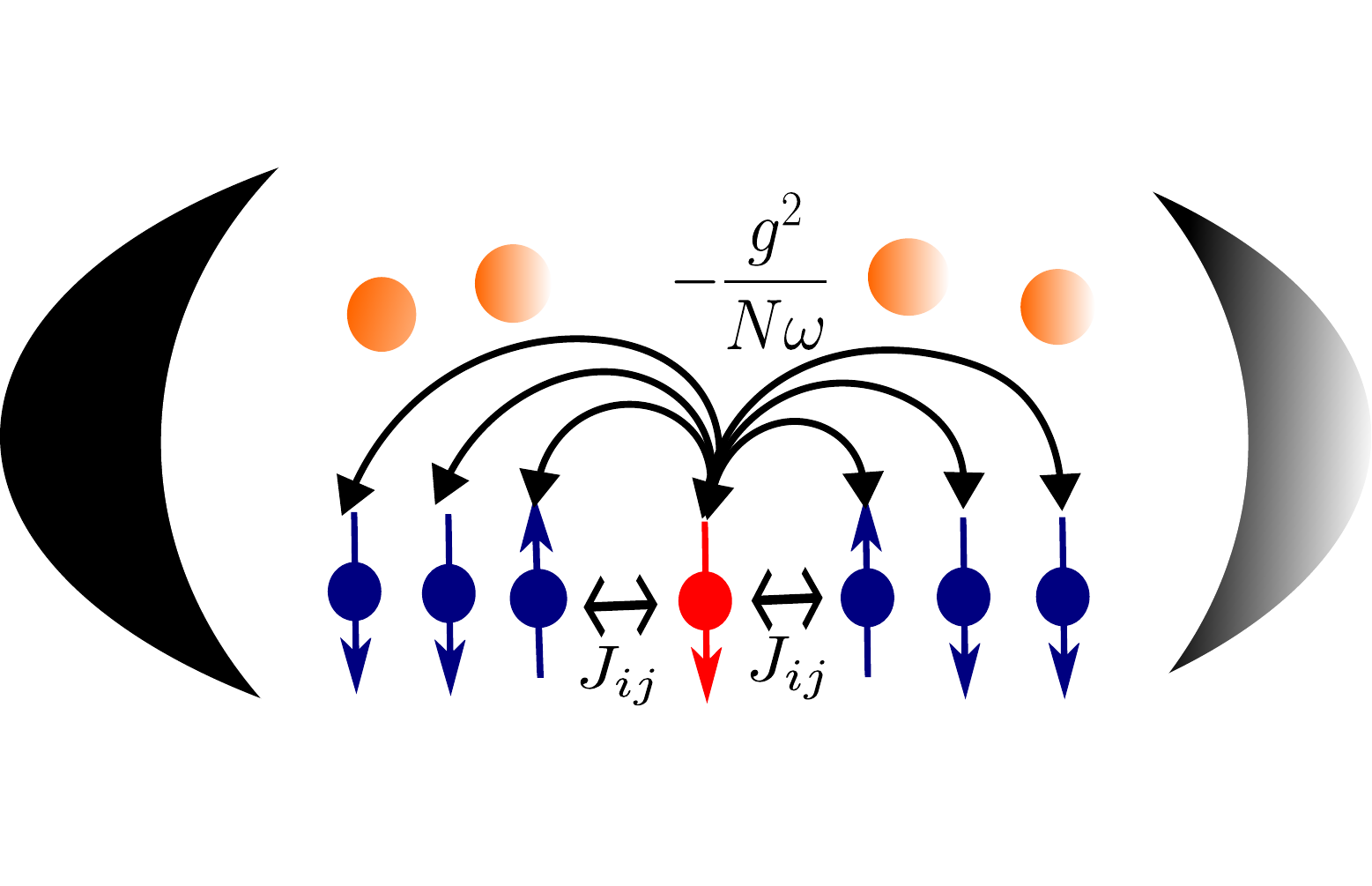} 
  \caption{When the spin system couples to a cavity photon, a long-range, all-to-all ferromagnetic interaction emerges between the spins. In the figure, a specific spin highlighted in red interacts with the other spins not only through the original internal spin-spin interactions $J_{ij}$, but also via a fully-connected intractions $-\frac{g^2}{N\omega}$ mediated by photons.}
  \label{fig:diagram2} 
\end{figure}

The second term is the fully connected part, the mean field solution known to become exact in the thermodynamic limit $N \to \infty$ \cite{sitter1937introduction}. 

The mean-field approximation is analytically equivalent to neglecting the square term of the fluctuation $\delta m_i = \sigma_i^{\mu} - m$ of the magnetization. In this case, the mean-field approximation for the all-to-all coupling term can be obtained as follows:
\begin{align}
  \sum_{i \neq j} \sigma_i^\mu \sigma_j^\mu \simeq -N^2m^2+2Nm\sum_{i=1}^N \sigma_i^\mu. \label{msigma}
\end{align}
We treat the fully connected interaction term within a mean-field framework. This approach formally neglects the collective fluctuation term $\dfrac{g^2}{N}\sum_{i \neq j} \delta m_i\delta m_j=\dfrac{g^2}{N}\sum_{i \neq j}(\sigma_i^\mu-m)(\sigma_j^\mu-m)$. However, as discussed in recent path-integral analyses \cite{PhysRevB.111.035156}, and also demonstrated via a displacement transformation in the supplemental material of Ref. \cite{Langheld2025}, the contribution of these macroscopic fluctuations scales as $O(1/N)$ and vanishes in the thermodynamic limit $N\to\infty$ . Therefore, rather than being a perturbative approximation limited to weak couplings, this mean-field treatment is asymptotically exact and well-justified for arbitrary light-matter coupling strengths in the macroscopic limit.

In the SRPT, a cavity electromagnetic field spontaneously emerges.
Therefore, it is essential to evaluate the expectation values of creation and annihilation operators of the cavity photons, as well as that of the photon number. 

We get
\begin{equation}
    \ev{a}=\ev{a^\dagger}=-\dfrac{gm}{\omega}\sqrt{N}.
\end{equation}
As can be seen from this, the emergence of a finite magnetization is accompanied by the development of a macroscopic electromagnetic field in the cavity \ref{sec:photon_stats}.

The expectation value of $\ev{a^\dagger a}$ satisfies
    \begin{align}
\ev{a^\dagger a} 
  = \dfrac{1}{e^{\beta\omega}-1} 
     + \frac{g^2}{N\omega^2} 
       \ev{\left( \sum_{i=1}^N \sigma_i^\mu \right)^2},
\end{align}
where we have taken into account the Bose-Einstein distribution $\ev{b^\dagger b}=\dfrac{1}{e^{\beta\omega}-1}$ \ref{sec:photon_stats}. Interestingly, the squared magnetization appears here as well.
By applying a partial mean-field approximation in the same manner as in the magnetization calculation, we obtain the following \ref{sec:photon_stats}:
\begin{equation}
    n  \simeq \dfrac{g^2m^2}{\omega^2}  = \left( \frac{\ev{a}}{\sqrt{N}} \right)^2. \label{n}
\end{equation}

\subsection{One-Dimensional Classical Spin Chain Coupled to a Cavity}

\begin{figure*}
    \centering
    \subfloat[]{\includegraphics[width=0.32\linewidth]{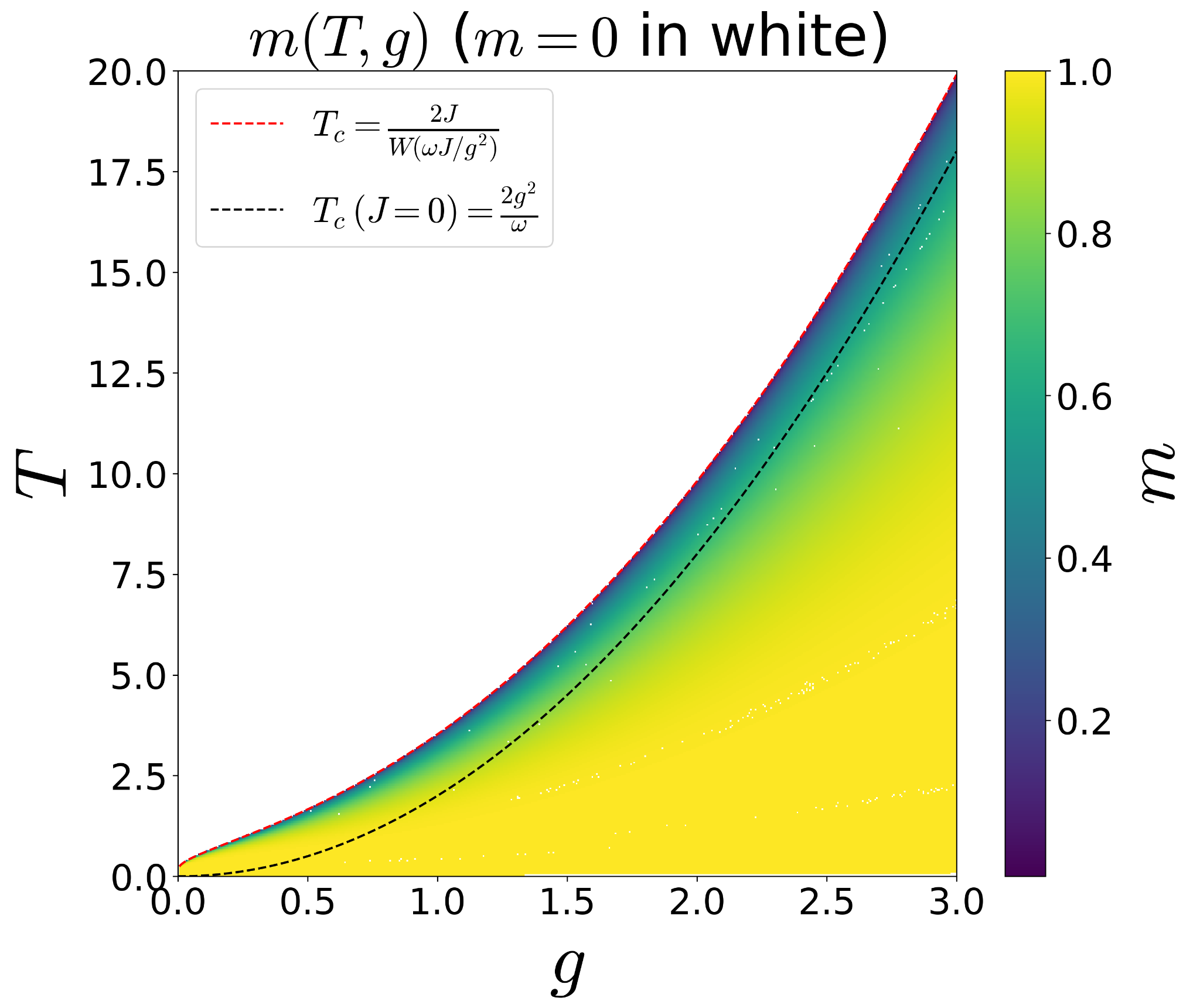}}\hfill
    \subfloat[]{\includegraphics[width=0.32\linewidth]{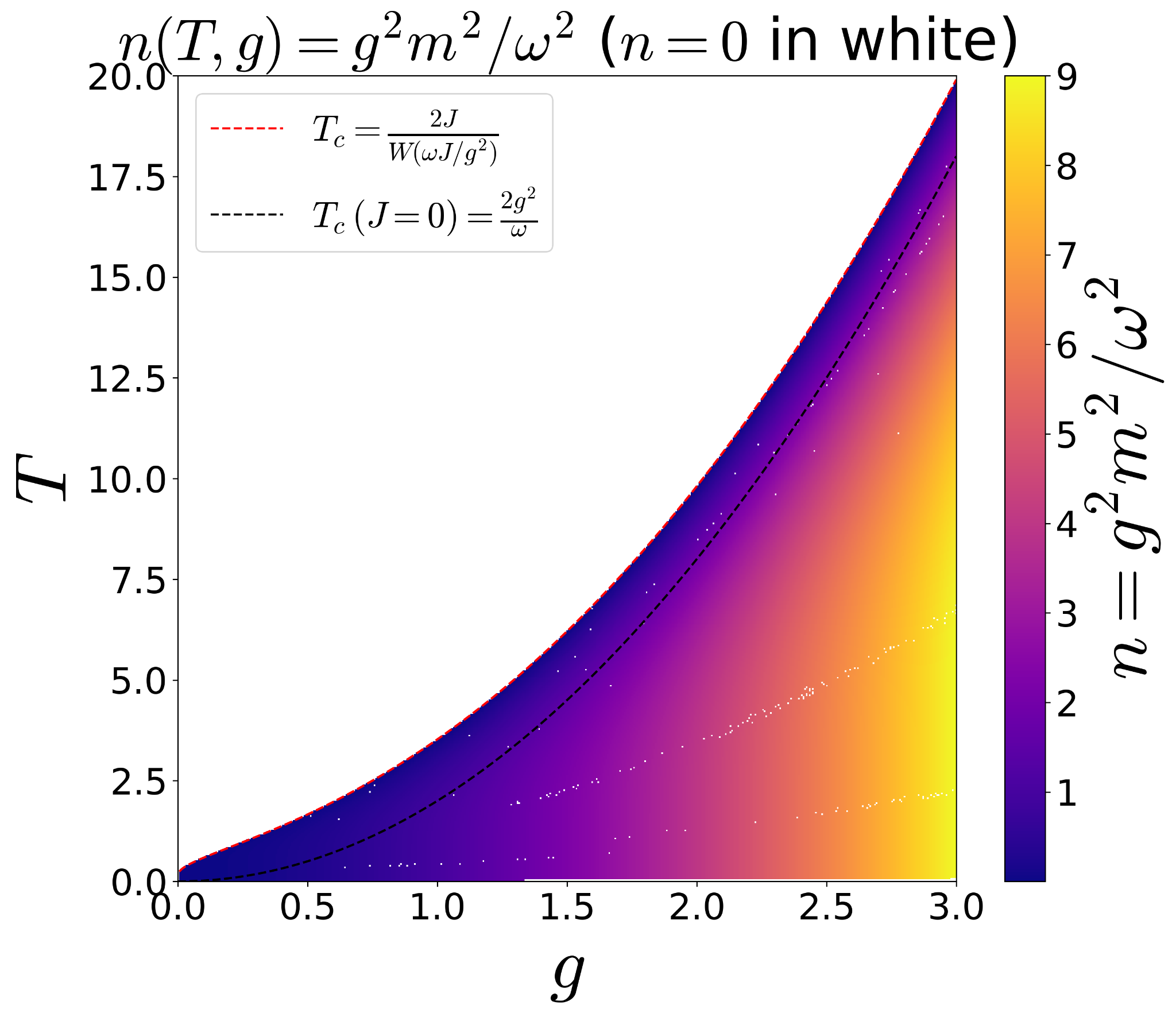}}\hfill
    \subfloat[]{\includegraphics[width=0.32\linewidth]{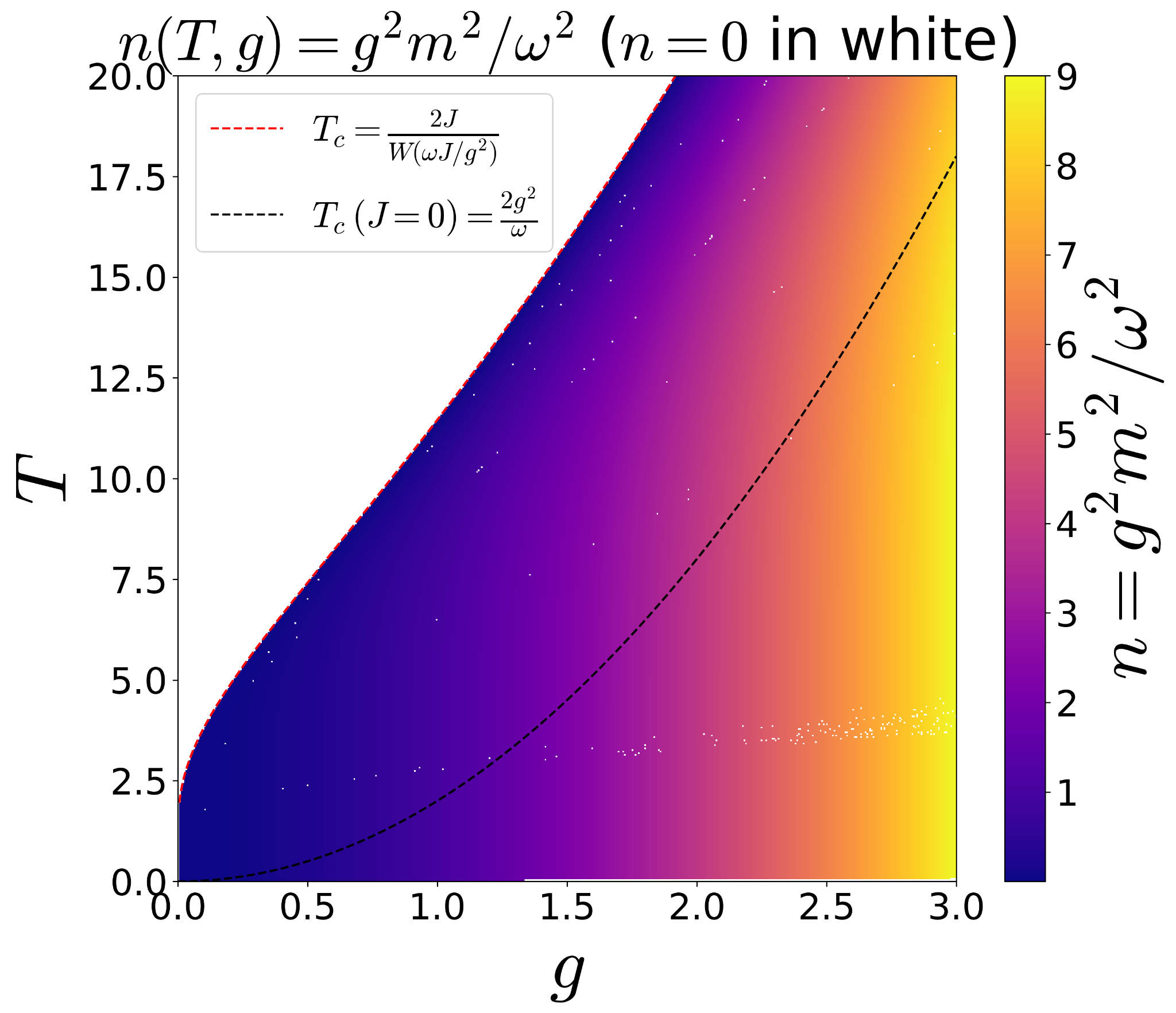}}\hfill
    \caption{(a) Magnetization $m$ and (b) normalized photon number $n = g^2 m^2 / \omega^2$ for a nearest-neighbor interaction $J=1$. 
(c) Normalized photon number $n$ for a stronger interaction, $J=10$. 
All quantities are plotted as a function of temperature $T$ and light--matter coupling strength $g$. The colored regions represent the ordered ferromagnetic, superradiant phase ($m, n \neq 0$), while the white region corresponds to the disordered paramagnetic phase ($m=n=0$). The red line is the exact critical temperature $T_c$ for the interacting chain, and the black line shows $T_c$ for the non-interacting ($J=0$) case for comparison. All calculations are performed with cavity frequency $\omega=1$.}
\label{fig:result}
\end{figure*}

As a specific example of a spin system for further calculation, let us consider a one-dimensional classical spin chain. This system is a specific example of the typical classical Ising model Eq.~(\ref{Ising}). The Hamiltonian is expressed as follows:
\begin{align}
  H_{\text{1D-Ising}}=-J\sum_{i=1}^N \sigma_i^z\sigma_{i+1}^z.
\end{align}
Here we take $J>0$ (ferromagnetic) and impose periodic boundary conditions, \(\sigma_{N+1}^z=\sigma_1^z\).
The exact solution for the magnetization when a classical external magnetic field $h$ is applied to the system is given by the following expression, where the Hamiltonian is defined as $H_{\text{1D-Ising}}-h\sum_{i=1}^N \sigma_i^z$ \cite{ising1925}:
\begin{align}
  m=\dfrac{\sinh(\beta h)}{\sqrt{(\sinh(\beta h))^2+e^{-4\beta J}}}. \label{magnetization}
\end{align}
In the following, we consider a model where this classical magnetic field is quantized as cavity photons. The Hamiltonian of the system is given by:
\begin{equation}
  H=-J\sum_{i=1}^N \sigma_i^z\sigma_{i+1}^z+\omega a^\dagger a+\frac{g}{\sqrt{N}}(a+a^\dagger)\sum_{i=1}^N \sigma_i^z.\label{model_1D}
\end{equation}
This corresponds to a specific example of the Hamiltonian given in Eq.~(\ref{model}) with $H_\text{spin}=H_{\text{1D-Ising}}$ and $\mu=z$.
When both the reversal of the spins $\sigma_i^z\to-\sigma_i^z$ of the entire system and the ``reversal of the photons'' $a\to -a$ are performed simultaneously, the Hamiltonian Eq.~(\ref{model_1D}) remains invariant. This is the $Z_2$ symmetry of the system.
By applying the same procedure as in the previous section, we diagonalize the Hamiltonian as follows.
\begin{align}
  H= -J \sum_{i=1}^N \sigma_i^z \sigma_{i+1}^z - \dfrac{g^2}{N\omega} \sum_{i \neq j} \sigma_i^z \sigma_j^z + \omega b^\dagger b + \text{const}.\label{1Dmodel}
\end{align}
Furthermore, applying the partial mean-field approximation results in the following effective Hamiltonian for the spins:
\begin{align}
  H_{\text{eff}}=-J \sum_{i=1}^N \sigma_i^z \sigma_{i+1}^z - \dfrac{2g^2m}{\omega} \sum_{i=1}^N \sigma_i^z .\label{1Dmodel_eff}
\end{align}
For this classical 1D model, the partial mean-field approximation becomes exact. It was pointed out in Baker's paper \cite{baker1963ising} that the partial mean-field approximation for this model is exact for calculating the transition point. Its rigorous derivation is presented in the appendix \ref{sec:ising_exact}.  The key feature of this appendix is the use of the saddle-point method, which yields results consistent with the approach by J. Román-Roch {\it et al.} based on the path integral representation of the action \cite{PhysRevB.111.035156}, as well as the exact mean-field decoupling demonstrated via a displacement transformation for the transverse-coupling case by Langheld et al. \cite{Langheld2025}. Furthermore, by explicitly evaluating the partition function via the Hubbard-Stratonovich transformation and saddle-point method, our derivation provides a complementary statistical mechanics perspective. The partition function of the effective spin Hamiltonian Eq.~(\ref{1Dmodel_eff}), which contains both short-range ($J$) and all-to-all interactions ($g$), can be rigorously evaluated. This involves applying a Hubbard-Stratonovich transformation \cite{stratonovich1957method,hubbard1959calculation} to the all-to-all term, followed by the method of steepest descents in the thermodynamic limit ($N \to \infty$). This mathematical procedure is identical to applying the partial mean-field approximation, thus proving its exactness. By interpreting $\frac{2g^2m}{\omega}$ as a classical external magnetic field, the system is effectively equivalent to a one-dimensional classical spin chain under an applied classical field. Therefore, utilizing the exact solution Eq.~\eqref{magnetization}, we obtain the following self-consistent equation:
\begin{equation}
    m = \frac{\sinh\left( 2\beta \dfrac{g^2}{\omega} m \right)}
    {\sqrt{\sinh^2\left( 2\beta \dfrac{g^2}{\omega} m \right) + e^{-4\beta J}}}. \label{self}
\end{equation}
This strategy of mapping the long-range interactions to a self-consistently determined effective field is a powerful approach that has been fruitfully employed in various contexts, such as coupled solid-state systems~\cite{Sandvik1999}, free fermions in a cavity~\cite{Keeling2014}, and other cavity QED platforms~\cite{Halati2017}.
From this self-consistent equation, we can obtain a second-order phase transition where the magnetization acquires a finite value in the low-temperature and strong-coupling regime. This second-order phase transition is consistent with the work of Baker \cite{baker1963ising}, which demonstrated that even a small amount of long-range interaction added to a short-range system forces the transition to be of the Bragg-Williams type.
The dependence of the order parameters $m$ and $n$ on the coupling strength $g$ is shown in Fig.~\ref{fig:result}.
The critical temperature $T_\text{c}$ can be determined exactly from the self-consistent equation \ref{sec:critical_temp}:
\begin{equation} \label{eq:Tc}
T_c=\frac{2g^2}{\omega}\exp{\left(\frac{2J}{T_c}\right)}=\frac{2J}{W(\omega J/g^2)},\quad 
T_c \xrightarrow[g\to\infty]{} \frac{2g^2}{\omega},
\end{equation}
where $W$ is the Lambert W function defined by $W(z)\,e^{W(z)}=z$. 
Equivalently, at fixed $T=\beta^{-1}$,
\begin{equation}
g_c(T)=\sqrt{\frac{\omega T}{2}}\;e^{-J/T}.
\end{equation}

The results shown in Fig.~\ref{fig:result} are obtained with $\omega$ = 1 fixed. The white region indicates the paramagnetic phase where the magnetization vanishes and the cavity field are zero.

As seen in Fig.~\ref{fig:result}, a ferromagnetic phase ($m\neq0$) accompanied by superradiant phase ($n\neq0$ or $\langle a \rangle\neq0$) appears in the strong-coupling, low-temperature regime, as indicated by the colored regions.

The red line represents the exact critical temperature within the regime discussed, while the black line corresponds to the critical temperature in the case of $J = 0$. The red and black lines coincide in the weak-coupling limit ($g\to0$). In the strong-coupling limit ($g\to\infty$), the red line asymptotically approaches the black line while not explicitly shown in the figure.

Moreover, from Eq.~\eqref{eq:Tc}, the dependence of the critical temperature on $J$ in the low-temperature region exhibits an exponential behavior. This suggests that, for the experimental detection of SRPT, not only the coupling strength $g$ with the cavity but also the strength $J$ of interactions within the material plays an important role.

\subsection{Comparison with Other Models}
The procedure shown above makes the picture of SRPT more clear even in the well-known Dicke model \cite{dicke1954}.
Applying a partial mean-field approximation to the spins in Dicke model yields the following form.
\begin{equation}
    H_{\text{Dicke-eff}} = \omega_0\sum_{i=1}^N \sigma_i^z -\dfrac{2g^2m^x}{\omega}\sum_{i=1}^N \sigma_i^x. \label{Dicke_mean}
\end{equation}
Here, $m^x$ denotes the magnetization in the $x$-direction. In this case as well, it shows a typical second-order phase transition. The critical coupling strength in this simple effective Hamiltonian can be exactly obtained as follows. 
\begin{equation}
    g_\text{c}= \sqrt{\dfrac{\omega\omega_0}{2}\coth{\beta\omega_0}} \underset{T=\beta^{-1} \to 0}{\longrightarrow} \sqrt{\dfrac{\omega\omega_0}{2}}. \label{g_c}
\end{equation}
This is consistent with what is shown in Ref.~\cite{Kirton2019}.
Moreover, Langheld {\it et al.} and Rohn {\it et al.} discussed the zero-temperature behavior of the following Hamiltonian \cite{Rohn:2020:IML, Langheld2025}, closely related with that in the present study: 
\begin{equation}
    H_\mathrm{T}=-J\sum_{i=1}^N \sigma_i^z\sigma_{i+1}^z+\omega a^\dagger a+\varepsilon\sum_{i=1}^N\sigma^z+\frac{g}{\sqrt{N}}(a+a^\dagger)\sum_{i=1}^N \sigma_i^x.
\end{equation}
Applying a partial mean-field approximation to spins in this Hamiltonian yields:
\begin{equation} \label{eq:Rohn}
    H_\text{T-eff} \underset{g \to \infty}{\longrightarrow} -\dfrac{2g^2m^x}{\omega}\sum_{i=1}^N \sigma_i^x.
\end{equation}
In the same manner as discussed above, in the limit where the coupling $g$ becomes large and the second term dominates, a second-order phase transition at finite temperature is expected to occur also in this Hamiltonian. However, numerical calculations at absolute zero temperature reported in Ref.~\cite{Rohn:2020:IML} have confirmed a first-order phase transition at $\varepsilon=0$, and recent quantum Monte Carlo simulations in one and two dimensions, alongside an exact one-dimensional solution provided in the appendix of Ref.~\cite{Langheld2025}, have investigated the effect of a finite longitudinal field $\varepsilon$, revealing that this first-order nature persists for small fields but gives way to a second-order transition via a multicritical point for larger fields.

Since this Hamiltonian possesses a \( Z_2 \times Z_2 \) symmetry \cite{Rohn:2020:IML}, which differs from that of our model, the difference may arise from the fact that the two systems belong to different universality classes. 

The second-order phase transition observed in our model at finite temperatures contrasts with the first-order transition found strictly at $T=0$~\cite{Rohn:2020:IML,Langheld2025}. However, just as a longitudinal field $\varepsilon$ melts the $T=0$ first-order transition into a second-order one~\cite{Langheld2025}, thermal fluctuations can play an analogous role. It is physically plausible that heating the system washes out the discontinuous nature of the $T=0$ transition, softening it into a continuous one. Exploring how the $T=0$ multicritical point extends into the finite-temperature phase diagram remains an interesting direction for future study.

\subsection{Conclusion}
The one-dimensional classical Ising chain is famously known to forbid a finite-temperature phase transition. The central finding of this work is that this 1D system does undergo a second-order magnetic phase transition when coupled to a cavity photon field. This transition is a realization of the SRPT, driven by a photon-mediated, all-to-all ferromagnetic interaction. The partial mean-field approximation, applied only to this emergent all-to-all term, is rigorously exact in the thermodynamic limit, despite the presence of the intrinsic 1D interaction ($J$). In the case of the classical 1D model, the rigorous procedure maps the entire problem onto the solvable 1D Ising chain in an effective longitudinal field, providing a new solvable SRPT. This mapping allows us to derive exact analytical expressions for the magnetization, critical temperature, and cavity field, providing the simplest solvable models of an SRPT coexisting with material interactions. 
The approach also serves to revalidate the transition in the well-known Dicke model. 
The mechanism of cavity-induced ordering may be observable in superconducting qubit arrays \cite{Zhang2014Quantum} and quasi-one-dimensional Ising-like magnetic materials (e.g.,$\text{CoNb}_2\text{O}_6$ \cite{Coldea2010}) coupled to cavity resonators.
In such materials, the magnetic transition temperature $T_N$ is typically limited by weak interchain couplings (e.g., $T_N = 2.95$~K for $\mathrm{CoNb_2O_6}$). 
Our exact solution demonstrates that the cavity-mediated all-to-all interaction induces a finite-temperature phase transition independent of the interchain couplings, indicating that the transition temperature can be enhanced by the cavity. 
To effectively couple with the spin dynamics, the cavity frequency $\omega$ should be on the same order as the intrachain exchange $J$. 
Since $J \sim 2$~meV in $\mathrm{CoNb_2O_6}$~\cite{Coldea2010}, the characteristic frequency is approximately $0.5$~THz, providing explicit motivation for the use of sub-terahertz cavity platforms. 
While dissipation is inherently present in experimental settings, the present study rigorously analyzes the exact finite-temperature thermodynamics as a fundamental benchmark in the absence of dissipation.

\vspace{1em}
\subsection*{Acknowledgement}
The authors are grateful for insightful discussions with Chisa Hotta and Tobias Asano. M. B. acknowledges support from the Research Foundation for Opto-Science and Technology and from the Japan Society for the Promotion of Science (JSPS) Grant Number JPJSJRP20221202 and KAKENHI Grant Numbers JP24K21526, JP25K00012, JP25K01691, and JP25K01694.

\bibliography{references}

\begin{thebibliography}{39}%
\makeatletter
\providecommand \@ifxundefined [1]{%
 \@ifx{#1\undefined}
}%
\providecommand \@ifnum [1]{%
 \ifnum #1\expandafter \@firstoftwo
 \else \expandafter \@secondoftwo
 \fi
}%
\providecommand \@ifx [1]{%
 \ifx #1\expandafter \@firstoftwo
 \else \expandafter \@secondoftwo
 \fi
}%
\providecommand \natexlab [1]{#1}%
\providecommand \enquote  [1]{``#1''}%
\providecommand \bibnamefont  [1]{#1}%
\providecommand \bibfnamefont [1]{#1}%
\providecommand \citenamefont [1]{#1}%
\providecommand \href@noop [0]{\@secondoftwo}%
\providecommand \href [0]{\begingroup \@sanitize@url \@href}%
\providecommand \@href[1]{\@@startlink{#1}\@@href}%
\providecommand \@@href[1]{\endgroup#1\@@endlink}%
\providecommand \@sanitize@url [0]{\catcode `\\12\catcode `\$12\catcode `\&12\catcode `\#12\catcode `\^12\catcode `\_12\catcode `\%12\relax}%
\providecommand \@@startlink[1]{}%
\providecommand \@@endlink[0]{}%
\providecommand \url  [0]{\begingroup\@sanitize@url \@url }%
\providecommand \@url [1]{\endgroup\@href {#1}{\urlprefix }}%
\providecommand \urlprefix  [0]{URL }%
\providecommand \Eprint [0]{\href }%
\providecommand \doibase [0]{https://doi.org/}%
\providecommand \selectlanguage [0]{\@gobble}%
\providecommand \bibinfo  [0]{\@secondoftwo}%
\providecommand \bibfield  [0]{\@secondoftwo}%
\providecommand \translation [1]{[#1]}%
\providecommand \BibitemOpen [0]{}%
\providecommand \bibitemStop [0]{}%
\providecommand \bibitemNoStop [0]{.\EOS\space}%
\providecommand \EOS [0]{\spacefactor3000\relax}%
\providecommand \BibitemShut  [1]{\csname bibitem#1\endcsname}%
\let\auto@bib@innerbib\@empty
\bibitem [{\citenamefont {Ising}(1925)}]{ising1925}%
  \BibitemOpen
  \bibfield  {author} {\bibinfo {author} {\bibfnamefont {E.}~\bibnamefont {Ising}},\ }\bibfield  {title} {\bibinfo {title} {Beitrag zur theorie des ferromagnetismus},\ }\href {https://doi.org/10.1007/BF02980577} {\bibfield  {journal} {\bibinfo  {journal} {Zeitschrift für Physik}\ }\textbf {\bibinfo {volume} {31}},\ \bibinfo {pages} {253} (\bibinfo {year} {1925})}\BibitemShut {NoStop}%
\bibitem [{\citenamefont {Forn-D{\'i}az}\ \emph {et~al.}(2019)\citenamefont {Forn-D{\'i}az}, \citenamefont {Lamata}, \citenamefont {Rico}, \citenamefont {Kono},\ and\ \citenamefont {Solano}}]{forn-diaz2019}%
  \BibitemOpen
  \bibfield  {author} {\bibinfo {author} {\bibfnamefont {P.}~\bibnamefont {Forn-D{\'i}az}}, \bibinfo {author} {\bibfnamefont {L.}~\bibnamefont {Lamata}}, \bibinfo {author} {\bibfnamefont {E.}~\bibnamefont {Rico}}, \bibinfo {author} {\bibfnamefont {J.}~\bibnamefont {Kono}},\ and\ \bibinfo {author} {\bibfnamefont {E.}~\bibnamefont {Solano}},\ }\bibfield  {title} {\bibinfo {title} {Ultrastrong coupling regimes of light--matter interaction},\ }\href {https://doi.org/10.1103/RevModPhys.91.025005} {\bibfield  {journal} {\bibinfo  {journal} {Reviews of Modern Physics}\ }\textbf {\bibinfo {volume} {91}},\ \bibinfo {pages} {025005} (\bibinfo {year} {2019})}\BibitemShut {NoStop}%
\bibitem [{\citenamefont {Kockum}\ \emph {et~al.}(2019)\citenamefont {Kockum}, \citenamefont {Miranowicz}, \citenamefont {Liberato}, \citenamefont {Savasta},\ and\ \citenamefont {Nori}}]{kockum2019}%
  \BibitemOpen
  \bibfield  {author} {\bibinfo {author} {\bibfnamefont {A.~F.}\ \bibnamefont {Kockum}}, \bibinfo {author} {\bibfnamefont {A.}~\bibnamefont {Miranowicz}}, \bibinfo {author} {\bibfnamefont {S.~D.}\ \bibnamefont {Liberato}}, \bibinfo {author} {\bibfnamefont {S.}~\bibnamefont {Savasta}},\ and\ \bibinfo {author} {\bibfnamefont {F.}~\bibnamefont {Nori}},\ }\bibfield  {title} {\bibinfo {title} {Ultrastrong coupling between light and matter},\ }\href {https://doi.org/10.1038/s42254-018-0006-2} {\bibfield  {journal} {\bibinfo  {journal} {Nature Reviews Physics}\ }\textbf {\bibinfo {volume} {1}},\ \bibinfo {pages} {19} (\bibinfo {year} {2019})}\BibitemShut {NoStop}%
\bibitem [{\citenamefont {Faust}\ and\ \citenamefont {Henry}(1966)}]{Faust1966-rm}%
  \BibitemOpen
  \bibfield  {author} {\bibinfo {author} {\bibfnamefont {W.~L.}\ \bibnamefont {Faust}}\ and\ \bibinfo {author} {\bibfnamefont {C.~H.}\ \bibnamefont {Henry}},\ }\bibfield  {title} {\bibinfo {title} {Mixing of visible and near-resonance infrared light in {GaP}},\ }\href {https://doi.org/10.1103/PhysRevLett.17.1265} {\bibfield  {journal} {\bibinfo  {journal} {Phys. Rev. Lett.}\ }\textbf {\bibinfo {volume} {17}},\ \bibinfo {pages} {1265} (\bibinfo {year} {1966})}\BibitemShut {NoStop}%
\bibitem [{\citenamefont {Zhang}\ \emph {et~al.}(2021)\citenamefont {Zhang}, \citenamefont {Hirori}, \citenamefont {Sekiguchi}, \citenamefont {Shimazaki}, \citenamefont {Iwasaki}, \citenamefont {Nakamura}, \citenamefont {Wakamiya},\ and\ \citenamefont {Kanemitsu}}]{Zhang2021-rw}%
  \BibitemOpen
  \bibfield  {author} {\bibinfo {author} {\bibfnamefont {Z.}~\bibnamefont {Zhang}}, \bibinfo {author} {\bibfnamefont {H.}~\bibnamefont {Hirori}}, \bibinfo {author} {\bibfnamefont {F.}~\bibnamefont {Sekiguchi}}, \bibinfo {author} {\bibfnamefont {A.}~\bibnamefont {Shimazaki}}, \bibinfo {author} {\bibfnamefont {Y.}~\bibnamefont {Iwasaki}}, \bibinfo {author} {\bibfnamefont {T.}~\bibnamefont {Nakamura}}, \bibinfo {author} {\bibfnamefont {A.}~\bibnamefont {Wakamiya}},\ and\ \bibinfo {author} {\bibfnamefont {Y.}~\bibnamefont {Kanemitsu}},\ }\bibfield  {title} {\bibinfo {title} {Ultrastrong coupling between {THz} phonons and photons caused by an enhanced vacuum electric field},\ }\href {https://doi.org/10.1103/PhysRevResearch.3.L032021} {\bibfield  {journal} {\bibinfo  {journal} {Phys. Rev. Res.}\ }\textbf {\bibinfo {volume} {3}},\ \bibinfo {pages} {L032021} (\bibinfo {year} {2021})}\BibitemShut {NoStop}%
\bibitem [{\citenamefont {Barra-Burillo}\ \emph {et~al.}(2021)\citenamefont {Barra-Burillo}, \citenamefont {Muniain}, \citenamefont {Catalano}, \citenamefont {Autore}, \citenamefont {Casanova}, \citenamefont {Hueso}, \citenamefont {Aizpurua}, \citenamefont {Esteban},\ and\ \citenamefont {Hillenbrand}}]{Barra-Burillo2021-hr}%
  \BibitemOpen
  \bibfield  {author} {\bibinfo {author} {\bibfnamefont {M.}~\bibnamefont {Barra-Burillo}}, \bibinfo {author} {\bibfnamefont {U.}~\bibnamefont {Muniain}}, \bibinfo {author} {\bibfnamefont {S.}~\bibnamefont {Catalano}}, \bibinfo {author} {\bibfnamefont {M.}~\bibnamefont {Autore}}, \bibinfo {author} {\bibfnamefont {F.}~\bibnamefont {Casanova}}, \bibinfo {author} {\bibfnamefont {L.~E.}\ \bibnamefont {Hueso}}, \bibinfo {author} {\bibfnamefont {J.}~\bibnamefont {Aizpurua}}, \bibinfo {author} {\bibfnamefont {R.}~\bibnamefont {Esteban}},\ and\ \bibinfo {author} {\bibfnamefont {R.}~\bibnamefont {Hillenbrand}},\ }\bibfield  {title} {\bibinfo {title} {Microcavity phonon polaritons from the weak to the ultrastrong phonon–photon coupling regime},\ }\href {https://doi.org/10.1038/s41467-021-26060-x} {\bibfield  {journal} {\bibinfo  {journal} {Nat. Commun.}\ }\textbf {\bibinfo {volume} {12}},\ \bibinfo {pages} {6206} (\bibinfo {year} {2021})}\BibitemShut {NoStop}%
\bibitem [{\citenamefont {Günter}\ \emph {et~al.}(2009)\citenamefont {Günter}, \citenamefont {Anappara}, \citenamefont {Hees}, \citenamefont {Sell}, \citenamefont {Biasiol}, \citenamefont {Sorba}, \citenamefont {De~Liberato}, \citenamefont {Ciuti}, \citenamefont {Tredicucci}, \citenamefont {Leitenstorfer},\ and\ \citenamefont {Huber}}]{Gunter2009-yd}%
  \BibitemOpen
  \bibfield  {author} {\bibinfo {author} {\bibfnamefont {G.}~\bibnamefont {Günter}}, \bibinfo {author} {\bibfnamefont {A.~A.}\ \bibnamefont {Anappara}}, \bibinfo {author} {\bibfnamefont {J.}~\bibnamefont {Hees}}, \bibinfo {author} {\bibfnamefont {A.}~\bibnamefont {Sell}}, \bibinfo {author} {\bibfnamefont {G.}~\bibnamefont {Biasiol}}, \bibinfo {author} {\bibfnamefont {L.}~\bibnamefont {Sorba}}, \bibinfo {author} {\bibfnamefont {S.}~\bibnamefont {De~Liberato}}, \bibinfo {author} {\bibfnamefont {C.}~\bibnamefont {Ciuti}}, \bibinfo {author} {\bibfnamefont {A.}~\bibnamefont {Tredicucci}}, \bibinfo {author} {\bibfnamefont {A.}~\bibnamefont {Leitenstorfer}},\ and\ \bibinfo {author} {\bibfnamefont {R.}~\bibnamefont {Huber}},\ }\bibfield  {title} {\bibinfo {title} {Sub-cycle switch-on of ultrastrong light-matter interaction},\ }\href {https://doi.org/10.1038/nature07838} {\bibfield  {journal} {\bibinfo  {journal} {Nature}\ }\textbf {\bibinfo {volume} {458}},\ \bibinfo {pages} {178} (\bibinfo {year}
  {2009})}\BibitemShut {NoStop}%
\bibitem [{\citenamefont {Niemczyk}\ \emph {et~al.}(2010)\citenamefont {Niemczyk}, \citenamefont {Deppe}, \citenamefont {Huebl}, \citenamefont {Menzel}, \citenamefont {Hocke}, \citenamefont {Schwarz}, \citenamefont {Garcia-Ripoll}, \citenamefont {Zueco}, \citenamefont {Hümmer}, \citenamefont {Solano}, \citenamefont {Marx},\ and\ \citenamefont {Gross}}]{Niemczyk2010-jk}%
  \BibitemOpen
  \bibfield  {author} {\bibinfo {author} {\bibfnamefont {T.}~\bibnamefont {Niemczyk}}, \bibinfo {author} {\bibfnamefont {F.}~\bibnamefont {Deppe}}, \bibinfo {author} {\bibfnamefont {H.}~\bibnamefont {Huebl}}, \bibinfo {author} {\bibfnamefont {E.~P.}\ \bibnamefont {Menzel}}, \bibinfo {author} {\bibfnamefont {F.}~\bibnamefont {Hocke}}, \bibinfo {author} {\bibfnamefont {M.~J.}\ \bibnamefont {Schwarz}}, \bibinfo {author} {\bibfnamefont {J.~J.}\ \bibnamefont {Garcia-Ripoll}}, \bibinfo {author} {\bibfnamefont {D.}~\bibnamefont {Zueco}}, \bibinfo {author} {\bibfnamefont {T.}~\bibnamefont {Hümmer}}, \bibinfo {author} {\bibfnamefont {E.}~\bibnamefont {Solano}}, \bibinfo {author} {\bibfnamefont {A.}~\bibnamefont {Marx}},\ and\ \bibinfo {author} {\bibfnamefont {R.}~\bibnamefont {Gross}},\ }\bibfield  {title} {\bibinfo {title} {Circuit quantum electrodynamics in the ultrastrong-coupling regime},\ }\href {https://doi.org/10.1038/nphys1730} {\bibfield  {journal} {\bibinfo  {journal} {Nat. Phys.}\ }\textbf {\bibinfo
  {volume} {6}},\ \bibinfo {pages} {772} (\bibinfo {year} {2010})},\ \Eprint {https://arxiv.org/abs/1003.2376} {1003.2376} \BibitemShut {NoStop}%
\bibitem [{\citenamefont {Yoshihara}\ \emph {et~al.}(2017)\citenamefont {Yoshihara}, \citenamefont {Fuse}, \citenamefont {Ashhab}, \citenamefont {Kakuyanagi}, \citenamefont {Saito},\ and\ \citenamefont {Semba}}]{Yoshihara2017-fa}%
  \BibitemOpen
  \bibfield  {author} {\bibinfo {author} {\bibfnamefont {F.}~\bibnamefont {Yoshihara}}, \bibinfo {author} {\bibfnamefont {T.}~\bibnamefont {Fuse}}, \bibinfo {author} {\bibfnamefont {S.}~\bibnamefont {Ashhab}}, \bibinfo {author} {\bibfnamefont {K.}~\bibnamefont {Kakuyanagi}}, \bibinfo {author} {\bibfnamefont {S.}~\bibnamefont {Saito}},\ and\ \bibinfo {author} {\bibfnamefont {K.}~\bibnamefont {Semba}},\ }\bibfield  {title} {\bibinfo {title} {Superconducting qubit-oscillator circuit beyond the ultrastrong-coupling regime},\ }\href {https://doi.org/10.1038/nphys3906} {\bibfield  {journal} {\bibinfo  {journal} {Nat. Phys.}\ }\textbf {\bibinfo {volume} {13}},\ \bibinfo {pages} {44} (\bibinfo {year} {2017})},\ \Eprint {https://arxiv.org/abs/1602.00415} {1602.00415} \BibitemShut {NoStop}%
\bibitem [{\citenamefont {Schwartz}\ \emph {et~al.}(2011)\citenamefont {Schwartz}, \citenamefont {Hutchison}, \citenamefont {Genet},\ and\ \citenamefont {Ebbesen}}]{Schwartz2011-ir}%
  \BibitemOpen
  \bibfield  {author} {\bibinfo {author} {\bibfnamefont {T.}~\bibnamefont {Schwartz}}, \bibinfo {author} {\bibfnamefont {J.~A.}\ \bibnamefont {Hutchison}}, \bibinfo {author} {\bibfnamefont {C.}~\bibnamefont {Genet}},\ and\ \bibinfo {author} {\bibfnamefont {T.~W.}\ \bibnamefont {Ebbesen}},\ }\bibfield  {title} {\bibinfo {title} {Reversible switching of ultrastrong light-molecule coupling},\ }\href {https://doi.org/10.1103/PhysRevLett.106.196405} {\bibfield  {journal} {\bibinfo  {journal} {Phys. Rev. Lett.}\ }\textbf {\bibinfo {volume} {106}},\ \bibinfo {pages} {196405} (\bibinfo {year} {2011})}\BibitemShut {NoStop}%
\bibitem [{\citenamefont {Scalari}\ \emph {et~al.}(2012)\citenamefont {Scalari}, \citenamefont {Maissen}, \citenamefont {Turcinková}, \citenamefont {Hagenmüller}, \citenamefont {De~Liberato}, \citenamefont {Ciuti}, \citenamefont {Reichl}, \citenamefont {Schuh}, \citenamefont {Wegscheider}, \citenamefont {Beck},\ and\ \citenamefont {Faist}}]{Scalari2012-rw}%
  \BibitemOpen
  \bibfield  {author} {\bibinfo {author} {\bibfnamefont {G.}~\bibnamefont {Scalari}}, \bibinfo {author} {\bibfnamefont {C.}~\bibnamefont {Maissen}}, \bibinfo {author} {\bibfnamefont {D.}~\bibnamefont {Turcinková}}, \bibinfo {author} {\bibfnamefont {D.}~\bibnamefont {Hagenmüller}}, \bibinfo {author} {\bibfnamefont {S.}~\bibnamefont {De~Liberato}}, \bibinfo {author} {\bibfnamefont {C.}~\bibnamefont {Ciuti}}, \bibinfo {author} {\bibfnamefont {C.}~\bibnamefont {Reichl}}, \bibinfo {author} {\bibfnamefont {D.}~\bibnamefont {Schuh}}, \bibinfo {author} {\bibfnamefont {W.}~\bibnamefont {Wegscheider}}, \bibinfo {author} {\bibfnamefont {M.}~\bibnamefont {Beck}},\ and\ \bibinfo {author} {\bibfnamefont {J.}~\bibnamefont {Faist}},\ }\bibfield  {title} {{\selectlanguage {english}\bibinfo {title} {Ultrastrong coupling of the cyclotron transition of a {2D} electron gas to a {THz} metamaterial}},\ }\href {https://doi.org/10.1126/science.1216022} {\bibfield  {journal} {\bibinfo  {journal} {Science}\ }\textbf {\bibinfo {volume}
  {335}},\ \bibinfo {pages} {1323} (\bibinfo {year} {2012})}\BibitemShut {NoStop}%
\bibitem [{\citenamefont {Bayer}\ \emph {et~al.}(2017)\citenamefont {Bayer}, \citenamefont {Pozimski}, \citenamefont {Schambeck}, \citenamefont {Schuh}, \citenamefont {Huber}, \citenamefont {Bougeard},\ and\ \citenamefont {Lange}}]{Bayer2017-kb}%
  \BibitemOpen
  \bibfield  {author} {\bibinfo {author} {\bibfnamefont {A.}~\bibnamefont {Bayer}}, \bibinfo {author} {\bibfnamefont {M.}~\bibnamefont {Pozimski}}, \bibinfo {author} {\bibfnamefont {S.}~\bibnamefont {Schambeck}}, \bibinfo {author} {\bibfnamefont {D.}~\bibnamefont {Schuh}}, \bibinfo {author} {\bibfnamefont {R.}~\bibnamefont {Huber}}, \bibinfo {author} {\bibfnamefont {D.}~\bibnamefont {Bougeard}},\ and\ \bibinfo {author} {\bibfnamefont {C.}~\bibnamefont {Lange}},\ }\bibfield  {title} {\bibinfo {title} {Terahertz light--matter interaction beyond unity coupling strength},\ }\href {https://doi.org/10.1021/acs.nanolett.7b03103} {\bibfield  {journal} {\bibinfo  {journal} {Nano Lett.}\ }\textbf {\bibinfo {volume} {17}},\ \bibinfo {pages} {6340} (\bibinfo {year} {2017})}\BibitemShut {NoStop}%
\bibitem [{\citenamefont {George}\ \emph {et~al.}(2016)\citenamefont {George}, \citenamefont {Chervy}, \citenamefont {Shalabney}, \citenamefont {Devaux}, \citenamefont {Hiura}, \citenamefont {Genet},\ and\ \citenamefont {Ebbesen}}]{George2016-kc}%
  \BibitemOpen
  \bibfield  {author} {\bibinfo {author} {\bibfnamefont {J.}~\bibnamefont {George}}, \bibinfo {author} {\bibfnamefont {T.}~\bibnamefont {Chervy}}, \bibinfo {author} {\bibfnamefont {A.}~\bibnamefont {Shalabney}}, \bibinfo {author} {\bibfnamefont {E.}~\bibnamefont {Devaux}}, \bibinfo {author} {\bibfnamefont {H.}~\bibnamefont {Hiura}}, \bibinfo {author} {\bibfnamefont {C.}~\bibnamefont {Genet}},\ and\ \bibinfo {author} {\bibfnamefont {T.~W.}\ \bibnamefont {Ebbesen}},\ }\bibfield  {title} {\bibinfo {title} {Multiple rabi splittings under ultrastrong vibrational coupling},\ }\href {https://doi.org/10.1103/PhysRevLett.117.153601} {\bibfield  {journal} {\bibinfo  {journal} {Phys. Rev. Lett.}\ }\textbf {\bibinfo {volume} {117}},\ \bibinfo {pages} {153601} (\bibinfo {year} {2016})},\ \Eprint {https://arxiv.org/abs/1609.01520} {1609.01520} \BibitemShut {NoStop}%
\bibitem [{\citenamefont {Mueller}\ \emph {et~al.}(2020)\citenamefont {Mueller}, \citenamefont {Okamura}, \citenamefont {Vieira}, \citenamefont {Juergensen}, \citenamefont {Lange}, \citenamefont {Barros}, \citenamefont {Schulz},\ and\ \citenamefont {Reich}}]{Mueller2020-bo}%
  \BibitemOpen
  \bibfield  {author} {\bibinfo {author} {\bibfnamefont {N.~S.}\ \bibnamefont {Mueller}}, \bibinfo {author} {\bibfnamefont {Y.}~\bibnamefont {Okamura}}, \bibinfo {author} {\bibfnamefont {B.~G.~M.}\ \bibnamefont {Vieira}}, \bibinfo {author} {\bibfnamefont {S.}~\bibnamefont {Juergensen}}, \bibinfo {author} {\bibfnamefont {H.}~\bibnamefont {Lange}}, \bibinfo {author} {\bibfnamefont {E.~B.}\ \bibnamefont {Barros}}, \bibinfo {author} {\bibfnamefont {F.}~\bibnamefont {Schulz}},\ and\ \bibinfo {author} {\bibfnamefont {S.}~\bibnamefont {Reich}},\ }\bibfield  {title} {\bibinfo {title} {Deep strong light–matter coupling in plasmonic nanoparticle crystals},\ }\href {https://doi.org/10.1038/s41586-020-2508-1} {\bibfield  {journal} {\bibinfo  {journal} {Nature}\ }\textbf {\bibinfo {volume} {583}},\ \bibinfo {pages} {780} (\bibinfo {year} {2020})}\BibitemShut {NoStop}%
\bibitem [{\citenamefont {Golovchanskiy}\ \emph {et~al.}(2021)\citenamefont {Golovchanskiy}, \citenamefont {Abramov}, \citenamefont {Stolyarov}, \citenamefont {Golubov}, \citenamefont {Kupriyanov}, \citenamefont {Ryazanov},\ and\ \citenamefont {Ustinov}}]{Golovchanskiy2021-hi}%
  \BibitemOpen
  \bibfield  {author} {\bibinfo {author} {\bibfnamefont {I.~A.}\ \bibnamefont {Golovchanskiy}}, \bibinfo {author} {\bibfnamefont {N.~N.}\ \bibnamefont {Abramov}}, \bibinfo {author} {\bibfnamefont {V.~S.}\ \bibnamefont {Stolyarov}}, \bibinfo {author} {\bibfnamefont {A.~A.}\ \bibnamefont {Golubov}}, \bibinfo {author} {\bibfnamefont {M.~Y.}\ \bibnamefont {Kupriyanov}}, \bibinfo {author} {\bibfnamefont {V.~V.}\ \bibnamefont {Ryazanov}},\ and\ \bibinfo {author} {\bibfnamefont {A.~V.}\ \bibnamefont {Ustinov}},\ }\bibfield  {title} {\bibinfo {title} {Approaching deep-strong on-chip photon-to-magnon coupling},\ }\href {https://doi.org/10.1103/PhysRevApplied.16.034029} {\bibfield  {journal} {\bibinfo  {journal} {Phys. Rev. Appl.}\ }\textbf {\bibinfo {volume} {16}},\ \bibinfo {pages} {034029} (\bibinfo {year} {2021})}\BibitemShut {NoStop}%
\bibitem [{\citenamefont {Nagarajan}\ \emph {et~al.}(2021)\citenamefont {Nagarajan}, \citenamefont {Thomas},\ and\ \citenamefont {Ebbesen}}]{Nagarajan2021-rd}%
  \BibitemOpen
  \bibfield  {author} {\bibinfo {author} {\bibfnamefont {K.}~\bibnamefont {Nagarajan}}, \bibinfo {author} {\bibfnamefont {A.}~\bibnamefont {Thomas}},\ and\ \bibinfo {author} {\bibfnamefont {T.~W.}\ \bibnamefont {Ebbesen}},\ }\bibfield  {title} {{\selectlanguage {english}\bibinfo {title} {Chemistry under vibrational strong coupling}},\ }\href {https://doi.org/10.1021/jacs.1c07420} {\bibfield  {journal} {\bibinfo  {journal} {J. Am. Chem. Soc.}\ }\textbf {\bibinfo {volume} {143}},\ \bibinfo {pages} {16877} (\bibinfo {year} {2021})}\BibitemShut {NoStop}%
\bibitem [{\citenamefont {Hepp}\ and\ \citenamefont {Lieb}(1973)}]{hepp1973}%
  \BibitemOpen
  \bibfield  {author} {\bibinfo {author} {\bibfnamefont {K.}~\bibnamefont {Hepp}}\ and\ \bibinfo {author} {\bibfnamefont {E.~H.}\ \bibnamefont {Lieb}},\ }\bibfield  {title} {\bibinfo {title} {On the superradiant phase transition for molecules in a quantized radiation field: The dicke maser model},\ }\href {https://doi.org/10.1016/0003-4916(73)90039-0} {\bibfield  {journal} {\bibinfo  {journal} {Annals of Physics}\ }\textbf {\bibinfo {volume} {76}},\ \bibinfo {pages} {360} (\bibinfo {year} {1973})}\BibitemShut {NoStop}%
\bibitem [{\citenamefont {Dicke}(1954)}]{dicke1954}%
  \BibitemOpen
  \bibfield  {author} {\bibinfo {author} {\bibfnamefont {R.~H.}\ \bibnamefont {Dicke}},\ }\bibfield  {title} {\bibinfo {title} {Coherence in spontaneous radiation processes},\ }\href {https://doi.org/10.1103/PhysRev.93.99} {\bibfield  {journal} {\bibinfo  {journal} {Physical Review}\ }\textbf {\bibinfo {volume} {93}},\ \bibinfo {pages} {99} (\bibinfo {year} {1954})}\BibitemShut {NoStop}%
\bibitem [{\citenamefont {Rohn}\ \emph {et~al.}(2020)\citenamefont {Rohn}, \citenamefont {H\"ormann}, \citenamefont {Genes},\ and\ \citenamefont {Schmidt}}]{Rohn:2020:IML}%
  \BibitemOpen
  \bibfield  {author} {\bibinfo {author} {\bibfnamefont {J.}~\bibnamefont {Rohn}}, \bibinfo {author} {\bibfnamefont {M.}~\bibnamefont {H\"ormann}}, \bibinfo {author} {\bibfnamefont {C.}~\bibnamefont {Genes}},\ and\ \bibinfo {author} {\bibfnamefont {K.~P.}\ \bibnamefont {Schmidt}},\ }\bibfield  {title} {\bibinfo {title} {Ising model in a light-induced quantized transverse field},\ }\href {https://doi.org/10.1103/PhysRevResearch.2.023131} {\bibfield  {journal} {\bibinfo  {journal} {Phys. Rev. Research}\ }\textbf {\bibinfo {volume} {2}},\ \bibinfo {pages} {023131} (\bibinfo {year} {2020})}\BibitemShut {NoStop}%
\bibitem [{\citenamefont {Nakamoto}\ \emph {et~al.}(2025)\citenamefont {Nakamoto}, \citenamefont {Takasan},\ and\ \citenamefont {Tsuji}}]{Nakamoto2025}%
  \BibitemOpen
  \bibfield  {author} {\bibinfo {author} {\bibfnamefont {T.}~\bibnamefont {Nakamoto}}, \bibinfo {author} {\bibfnamefont {K.}~\bibnamefont {Takasan}},\ and\ \bibinfo {author} {\bibfnamefont {N.}~\bibnamefont {Tsuji}},\ }\bibfield  {title} {\bibinfo {title} {One-dimensional extended hubbard model coupled with an optical cavity},\ }\href {https://doi.org/https://doi.org/10.1103/pxkp-trx5} {\bibfield  {journal} {\bibinfo  {journal} {Phys. Rev. B}\ }\textbf {\bibinfo {volume} {112}},\ \bibinfo {pages} {155150} (\bibinfo {year} {2025})}\BibitemShut {NoStop}%
\bibitem [{\citenamefont {Langheld}\ \emph {et~al.}(2025)\citenamefont {Langheld}, \citenamefont {H{\"o}rmann},\ and\ \citenamefont {Schmidt}}]{Langheld2025}%
  \BibitemOpen
  \bibfield  {author} {\bibinfo {author} {\bibfnamefont {A.}~\bibnamefont {Langheld}}, \bibinfo {author} {\bibfnamefont {M.}~\bibnamefont {H{\"o}rmann}},\ and\ \bibinfo {author} {\bibfnamefont {K.~P.}\ \bibnamefont {Schmidt}},\ }\bibfield  {title} {\bibinfo {title} {Quantum phase diagrams of dicke-ising models by a wormhole algorithm},\ }\href {https://doi.org/https://doi.org/10.1103/lcvj-ksct} {\bibfield  {journal} {\bibinfo  {journal} {Phys. Rev. B}\ }\textbf {\bibinfo {volume} {112}},\ \bibinfo {pages} {L161123} (\bibinfo {year} {2025})}\BibitemShut {NoStop}%
\bibitem [{\citenamefont {Shapiro}\ \emph {et~al.}(2025)\citenamefont {Shapiro}, \citenamefont {Weber}, \citenamefont {Bode}, \citenamefont {Wilhelm},\ and\ \citenamefont {Bagrets}}]{Shapiro2025}%
  \BibitemOpen
  \bibfield  {author} {\bibinfo {author} {\bibfnamefont {D.~S.}\ \bibnamefont {Shapiro}}, \bibinfo {author} {\bibfnamefont {Y.}~\bibnamefont {Weber}}, \bibinfo {author} {\bibfnamefont {T.}~\bibnamefont {Bode}}, \bibinfo {author} {\bibfnamefont {F.~K.}\ \bibnamefont {Wilhelm}},\ and\ \bibinfo {author} {\bibfnamefont {D.}~\bibnamefont {Bagrets}},\ }\bibfield  {title} {\bibinfo {title} {Digital-analog simulations of {S}chr{\"o}dinger cat states in the {D}icke-{I}sing model},\ }\href {https://doi.org/https://doi.org/10.1103/wbp6-y3vd} {\bibfield  {journal} {\bibinfo  {journal} {Phys. Rev. A}\ }\textbf {\bibinfo {volume} {112}},\ \bibinfo {pages} {042412} (\bibinfo {year} {2025})}\BibitemShut {NoStop}%
\bibitem [{\citenamefont {Koziol}\ \emph {et~al.}(2025)\citenamefont {Koziol}, \citenamefont {Langheld},\ and\ \citenamefont {Schmidt}}]{Koziol2025}%
  \BibitemOpen
  \bibfield  {author} {\bibinfo {author} {\bibfnamefont {J.~A.}\ \bibnamefont {Koziol}}, \bibinfo {author} {\bibfnamefont {A.}~\bibnamefont {Langheld}},\ and\ \bibinfo {author} {\bibfnamefont {K.~P.}\ \bibnamefont {Schmidt}},\ }\bibfield  {title} {\bibinfo {title} {Melting of devil's staircases in the long-range {D}icke-{I}sing model},\ }\href {https://doi.org/https://doi.org/10.1103/syps-9r7r} {\bibfield  {journal} {\bibinfo  {journal} {Phys. Rev. B}\ }\textbf {\bibinfo {volume} {111}},\ \bibinfo {pages} {224427} (\bibinfo {year} {2025})}\BibitemShut {NoStop}%
\bibitem [{\citenamefont {Cortese}\ \emph {et~al.}(2017)\citenamefont {Cortese}, \citenamefont {Garziano},\ and\ \citenamefont {De~Liberato}}]{Cortese2017}%
  \BibitemOpen
  \bibfield  {author} {\bibinfo {author} {\bibfnamefont {E.}~\bibnamefont {Cortese}}, \bibinfo {author} {\bibfnamefont {L.}~\bibnamefont {Garziano}},\ and\ \bibinfo {author} {\bibfnamefont {S.}~\bibnamefont {De~Liberato}},\ }\bibfield  {title} {\bibinfo {title} {Polariton spectrum of the dicke-ising model},\ }\href {https://doi.org/10.1103/PhysRevA.96.053861} {\bibfield  {journal} {\bibinfo  {journal} {Phys. Rev. A}\ }\textbf {\bibinfo {volume} {96}},\ \bibinfo {pages} {053861} (\bibinfo {year} {2017})}\BibitemShut {NoStop}%
\bibitem [{\citenamefont {Lee}\ and\ \citenamefont {Johnson}(2004)}]{Lee2004}%
  \BibitemOpen
  \bibfield  {author} {\bibinfo {author} {\bibfnamefont {C.~F.}\ \bibnamefont {Lee}}\ and\ \bibinfo {author} {\bibfnamefont {N.~F.}\ \bibnamefont {Johnson}},\ }\bibfield  {title} {\bibinfo {title} {First-order superradiant phase transitions in a multiqubit cavity system},\ }\href {https://doi.org/10.1103/PhysRevLett.93.083001} {\bibfield  {journal} {\bibinfo  {journal} {Phys. Rev. Lett.}\ }\textbf {\bibinfo {volume} {93}},\ \bibinfo {pages} {083001} (\bibinfo {year} {2004})}\BibitemShut {NoStop}%
\bibitem [{\citenamefont {Schellenberger}\ and\ \citenamefont {Schmidt}(2024)}]{Schellenberger2024}%
  \BibitemOpen
  \bibfield  {author} {\bibinfo {author} {\bibfnamefont {A.}~\bibnamefont {Schellenberger}}\ and\ \bibinfo {author} {\bibfnamefont {K.~P.}\ \bibnamefont {Schmidt}},\ }\bibfield  {title} {\bibinfo {title} {(almost) everything is a {Dicke} model - mapping non-superradiant correlated light-matter systems to the exactly solvable {Dicke} model},\ }\href {https://doi.org/10.21468/SciPostPhysCore.7.3.038} {\bibfield  {journal} {\bibinfo  {journal} {SciPost Phys. Core}\ }\textbf {\bibinfo {volume} {7}},\ \bibinfo {pages} {038} (\bibinfo {year} {2024})}\BibitemShut {NoStop}%
\bibitem [{\citenamefont {Rom\'an-Roche}\ \emph {et~al.}(2025{\natexlab{a}})\citenamefont {Rom\'an-Roche}, \citenamefont {G\'omez-Le\'on}, \citenamefont {Luis},\ and\ \citenamefont {Zueco}}]{RomanRoche2025Bound}%
  \BibitemOpen
  \bibfield  {author} {\bibinfo {author} {\bibfnamefont {J.}~\bibnamefont {Rom\'an-Roche}}, \bibinfo {author} {\bibfnamefont {A.}~\bibnamefont {G\'omez-Le\'on}}, \bibinfo {author} {\bibfnamefont {F.}~\bibnamefont {Luis}},\ and\ \bibinfo {author} {\bibfnamefont {D.}~\bibnamefont {Zueco}},\ }\bibfield  {title} {\bibinfo {title} {Bound polariton states in the {Dicke-Ising} model},\ }\href {https://doi.org/10.1515/nanoph-2024-0568} {\bibfield  {journal} {\bibinfo  {journal} {Nanophotonics}\ }\textbf {\bibinfo {volume} {14}},\ \bibinfo {pages} {2053} (\bibinfo {year} {2025}{\natexlab{a}})}\BibitemShut {NoStop}%
\bibitem [{\citenamefont {Mendon\c{c}a}\ \emph {et~al.}(2025)\citenamefont {Mendon\c{c}a}, \citenamefont {Jachymski},\ and\ \citenamefont {Wang}}]{Mendonca2025}%
  \BibitemOpen
  \bibfield  {author} {\bibinfo {author} {\bibfnamefont {J.~P.}\ \bibnamefont {Mendon\c{c}a}}, \bibinfo {author} {\bibfnamefont {K.}~\bibnamefont {Jachymski}},\ and\ \bibinfo {author} {\bibfnamefont {Y.}~\bibnamefont {Wang}},\ }\bibfield  {title} {\bibinfo {title} {Role of matter interactions in superradiant phenomena},\ }\href@noop {} {\bibfield  {journal} {\bibinfo  {journal} {Phys. Rev. Lett.}\ }\textbf {\bibinfo {volume} {135}},\ \bibinfo {pages} {133601} (\bibinfo {year} {2025})},\ \bibinfo {note} {currently under appeal, see Comment by M. H\"ormann, A. Langheld, J. Leibig, A. Schellenberger, and K. P. Schmidt, arXiv:2511.08452}\BibitemShut {NoStop}%
\bibitem [{\citenamefont {Kirton}\ \emph {et~al.}(2019)\citenamefont {Kirton}, \citenamefont {Roses}, \citenamefont {Keeling},\ and\ \citenamefont {Dalla~Torre}}]{Kirton2019}%
  \BibitemOpen
  \bibfield  {author} {\bibinfo {author} {\bibfnamefont {P.}~\bibnamefont {Kirton}}, \bibinfo {author} {\bibfnamefont {M.~M.}\ \bibnamefont {Roses}}, \bibinfo {author} {\bibfnamefont {J.}~\bibnamefont {Keeling}},\ and\ \bibinfo {author} {\bibfnamefont {E.~G.}\ \bibnamefont {Dalla~Torre}},\ }\bibfield  {title} {\bibinfo {title} {{Introduction to the Dicke model: from equilibrium to nonequilibrium, and vice versa}},\ }\href {https://doi.org/10.1002/qute.201800043} {\bibfield  {journal} {\bibinfo  {journal} {Advanced Quantum Technologies}\ }\textbf {\bibinfo {volume} {2}},\ \bibinfo {pages} {1800043} (\bibinfo {year} {2019})}\BibitemShut {NoStop}%
\bibitem [{\citenamefont {Sitter}(1937)}]{sitter1937introduction}%
  \BibitemOpen
  \bibfield  {author} {\bibinfo {author} {\bibfnamefont {F.}~\bibnamefont {Sitter}},\ }\href@noop {} {\emph {\bibinfo {title} {Introduction to Ferromagnetism}}}\ (\bibinfo  {publisher} {McGraw-Hill Book Company, Inc.},\ \bibinfo {address} {New York},\ \bibinfo {year} {1937})\BibitemShut {NoStop}%
\bibitem [{\citenamefont {Rom\'an-Roche}\ \emph {et~al.}(2025{\natexlab{b}})\citenamefont {Rom\'an-Roche}, \citenamefont {G\'omez-Le\'on}, \citenamefont {Luis},\ and\ \citenamefont {Zueco}}]{PhysRevB.111.035156}%
  \BibitemOpen
  \bibfield  {author} {\bibinfo {author} {\bibfnamefont {J.}~\bibnamefont {Rom\'an-Roche}}, \bibinfo {author} {\bibfnamefont {A.}~\bibnamefont {G\'omez-Le\'on}}, \bibinfo {author} {\bibfnamefont {F.}~\bibnamefont {Luis}},\ and\ \bibinfo {author} {\bibfnamefont {D.}~\bibnamefont {Zueco}},\ }\bibfield  {title} {\bibinfo {title} {Linear response theory for cavity qed materials at arbitrary light-matter coupling strengths},\ }\href {https://doi.org/10.1103/PhysRevB.111.035156} {\bibfield  {journal} {\bibinfo  {journal} {Phys. Rev. B}\ }\textbf {\bibinfo {volume} {111}},\ \bibinfo {pages} {035156} (\bibinfo {year} {2025}{\natexlab{b}})}\BibitemShut {NoStop}%
\bibitem [{\citenamefont {Baker}(1963)}]{baker1963ising}%
  \BibitemOpen
  \bibfield  {author} {\bibinfo {author} {\bibfnamefont {G.~A.}\ \bibnamefont {Baker}, \bibfnamefont {Jr.}},\ }\bibfield  {title} {\bibinfo {title} {Ising model with a long-range interaction in the presence of residual short-range interactions},\ }\href {https://doi.org/10.1103/PhysRev.130.1406} {\bibfield  {journal} {\bibinfo  {journal} {Physical Review}\ }\textbf {\bibinfo {volume} {130}},\ \bibinfo {pages} {1406} (\bibinfo {year} {1963})}\BibitemShut {NoStop}%
\bibitem [{\citenamefont {Stratonovich}(1957)}]{stratonovich1957method}%
  \BibitemOpen
  \bibfield  {author} {\bibinfo {author} {\bibfnamefont {R.~L.}\ \bibnamefont {Stratonovich}},\ }\bibfield  {title} {\bibinfo {title} {On a method of calculating quantum distribution functions},\ }\href@noop {} {\bibfield  {journal} {\bibinfo  {journal} {Doklady Akademii Nauk SSSR}\ }\textbf {\bibinfo {volume} {115}},\ \bibinfo {pages} {1097} (\bibinfo {year} {1957})}\BibitemShut {NoStop}%
\bibitem [{\citenamefont {Hubbard}(1959)}]{hubbard1959calculation}%
  \BibitemOpen
  \bibfield  {author} {\bibinfo {author} {\bibfnamefont {J.}~\bibnamefont {Hubbard}},\ }\bibfield  {title} {\bibinfo {title} {Calculation of partition functions},\ }\href {https://doi.org/10.1103/PhysRevLett.3.77} {\bibfield  {journal} {\bibinfo  {journal} {Physical Review Letters}\ }\textbf {\bibinfo {volume} {3}},\ \bibinfo {pages} {77} (\bibinfo {year} {1959})}\BibitemShut {NoStop}%
\bibitem [{\citenamefont {Sandvik}(1999)}]{Sandvik1999}%
  \BibitemOpen
  \bibfield  {author} {\bibinfo {author} {\bibfnamefont {A.~W.}\ \bibnamefont {Sandvik}},\ }\bibfield  {title} {\bibinfo {title} {Multichain mean-field theory of quasi-one-dimensional quantum spin systems},\ }\href {https://doi.org/10.1103/PhysRevLett.83.3069} {\bibfield  {journal} {\bibinfo  {journal} {Phys. Rev. Lett.}\ }\textbf {\bibinfo {volume} {83}},\ \bibinfo {pages} {3069} (\bibinfo {year} {1999})}\BibitemShut {NoStop}%
\bibitem [{\citenamefont {Keeling}\ \emph {et~al.}(2014)\citenamefont {Keeling}, \citenamefont {Bhaseen},\ and\ \citenamefont {Simons}}]{Keeling2014}%
  \BibitemOpen
  \bibfield  {author} {\bibinfo {author} {\bibfnamefont {J.}~\bibnamefont {Keeling}}, \bibinfo {author} {\bibfnamefont {M.~J.}\ \bibnamefont {Bhaseen}},\ and\ \bibinfo {author} {\bibfnamefont {B.~D.}\ \bibnamefont {Simons}},\ }\bibfield  {title} {\bibinfo {title} {Fermionic superradiance in a transversely pumped optical cavity},\ }\href {https://doi.org/10.1103/PhysRevLett.112.143002} {\bibfield  {journal} {\bibinfo  {journal} {Phys. Rev. Lett.}\ }\textbf {\bibinfo {volume} {112}},\ \bibinfo {pages} {143002} (\bibinfo {year} {2014})}\BibitemShut {NoStop}%
\bibitem [{\citenamefont {Halati}\ \emph {et~al.}(2017)\citenamefont {Halati}, \citenamefont {Sheikhan},\ and\ \citenamefont {Kollath}}]{Halati2017}%
  \BibitemOpen
  \bibfield  {author} {\bibinfo {author} {\bibfnamefont {C.-M.}\ \bibnamefont {Halati}}, \bibinfo {author} {\bibfnamefont {A.}~\bibnamefont {Sheikhan}},\ and\ \bibinfo {author} {\bibfnamefont {C.}~\bibnamefont {Kollath}},\ }\bibfield  {title} {\bibinfo {title} {Cavity-induced artificial gauge field in a bose-hubbard ladder},\ }\href {https://doi.org/10.1103/PhysRevA.96.063621} {\bibfield  {journal} {\bibinfo  {journal} {Phys. Rev. A}\ }\textbf {\bibinfo {volume} {96}},\ \bibinfo {pages} {063621} (\bibinfo {year} {2017})}\BibitemShut {NoStop}%
\bibitem [{\citenamefont {Zhang}\ \emph {et~al.}(2014)\citenamefont {Zhang}, \citenamefont {Yu}, \citenamefont {Liang}, \citenamefont {Chen}, \citenamefont {Jia},\ and\ \citenamefont {Nori}}]{Zhang2014Quantum}%
  \BibitemOpen
  \bibfield  {author} {\bibinfo {author} {\bibfnamefont {Y.-Q.}\ \bibnamefont {Zhang}}, \bibinfo {author} {\bibfnamefont {L.-H.}\ \bibnamefont {Yu}}, \bibinfo {author} {\bibfnamefont {J.-Q.}\ \bibnamefont {Liang}}, \bibinfo {author} {\bibfnamefont {G.}~\bibnamefont {Chen}}, \bibinfo {author} {\bibfnamefont {S.}~\bibnamefont {Jia}},\ and\ \bibinfo {author} {\bibfnamefont {F.}~\bibnamefont {Nori}},\ }\bibfield  {title} {\bibinfo {title} {Quantum phases in circuit qed with a superconducting qubit array},\ }\href {https://doi.org/10.1038/srep04083} {\bibfield  {journal} {\bibinfo  {journal} {Scientific Reports}\ }\textbf {\bibinfo {volume} {4}},\ \bibinfo {pages} {4083} (\bibinfo {year} {2014})}\BibitemShut {NoStop}%
\bibitem [{\citenamefont {Coldea}\ \emph {et~al.}(2010)\citenamefont {Coldea}, \citenamefont {Tennant}, \citenamefont {Wheeler}, \citenamefont {Wawrzynska}, \citenamefont {Prabhakaran}, \citenamefont {Telling}, \citenamefont {Habicht}, \citenamefont {Smeibidl},\ and\ \citenamefont {Kiefer}}]{Coldea2010}%
  \BibitemOpen
  \bibfield  {author} {\bibinfo {author} {\bibfnamefont {R.}~\bibnamefont {Coldea}}, \bibinfo {author} {\bibfnamefont {D.~A.}\ \bibnamefont {Tennant}}, \bibinfo {author} {\bibfnamefont {E.~M.}\ \bibnamefont {Wheeler}}, \bibinfo {author} {\bibfnamefont {E.}~\bibnamefont {Wawrzynska}}, \bibinfo {author} {\bibfnamefont {D.}~\bibnamefont {Prabhakaran}}, \bibinfo {author} {\bibfnamefont {M.}~\bibnamefont {Telling}}, \bibinfo {author} {\bibfnamefont {K.}~\bibnamefont {Habicht}}, \bibinfo {author} {\bibfnamefont {P.}~\bibnamefont {Smeibidl}},\ and\ \bibinfo {author} {\bibfnamefont {K.}~\bibnamefont {Kiefer}},\ }\bibfield  {title} {\bibinfo {title} {Quantum criticality in an ising chain: Experimental evidence for emergent e8 symmetry},\ }\href {https://doi.org/10.1126/science.1180085} {\bibfield  {journal} {\bibinfo  {journal} {Science}\ }\textbf {\bibinfo {volume} {327}},\ \bibinfo {pages} {177} (\bibinfo {year} {2010})}\BibitemShut {NoStop}%
\end{thebibliography}%

\appendix

\onecolumngrid

\section{General Theory: Separation of Variables via Polaron Transformation}
\label{sec:general_theory}

We consider a general spin Hamiltonian $H_{\text{spin}}$ coupled to a single cavity photon mode via a specific spin component $\sigma^{\mu}$ (where $\mu = x, y, \text{or } z$). The total Hamiltonian is given by Eq.~(\ref{model}) in the main text:
\begin{equation}
    H = H_{\text{spin}} + \omega a^{\dagger}a + \frac{g}{\sqrt{N}}(a+a^{\dagger})\sum_{i=1}^{N}\sigma_{i}^{\mu}
    \label{eq:sm_h_general}
\end{equation}
Our goal is to separate the photon degrees of freedom from the spin system to derive an effective spin model.

\subsection{Polaron Transformation and Effective Hamiltonian}
We introduce a "displaced" bosonic operator $b$ by completing the square for the photon field. This corresponds to a unitary polaron transformation (or displacement transformation):
\begin{equation}
    b = a + \frac{g}{\omega\sqrt{N}}\sum_{i=1}^{N}\sigma_{i}^{\mu}.
    \label{eq:sm_b_def}
\end{equation}
It is straightforward to verify that $b$ satisfies the canonical bosonic commutation relation $[b, b^\dagger] = [a, a^\dagger] = 1$.
Substituting $a = b - \frac{g}{\omega\sqrt{N}}\sum_{i}\sigma_{i}^{\mu}$ into Eq.~(\ref{eq:sm_h_general}), the total Hamiltonian transforms into:
\begin{equation}
    H = H_{\text{eff}} + \omega b^{\dagger}b, \quad \text{with} \quad H_{\text{eff}} = H_{\text{spin}} - \frac{g^{2}}{N\omega}\left(\sum_{i=1}^{N}\sigma_{i}^{\mu}\right)^{2}
    \label{eq:sm_h_separated}
\end{equation}
Here, $H_{\text{eff}}$ describes the spins with an induced infinite-range interaction. The term $\omega b^\dagger b$ represents a free boson.

\subsection{Rigorous Justification for Decoupling}
To treat the spins and photons as statistically independent systems, we must show that the partition function factorizes. This can be proven rigorously using the properties of the trace, without relying on specific commutation relations between $b$ and the internal spin dynamics.

Let $D$ be the unitary displacement operator that generates the transformation in Eq.~(\ref{eq:sm_b_def}):
\begin{equation}
    D = \exp\left[ \frac{g}{\omega\sqrt{N}}\left(\sum_{i}\sigma_{i}^{\mu}\right)(a^{\dagger} - a) \right], \quad \text{such that} \quad b = D^{\dagger} a D.
\end{equation}
The Hamiltonian can be expressed as $H = D^{\dagger} (H_{\text{eff}} + \omega a^{\dagger}a) D$. The partition function $Z$ is given by:
\begin{equation}
    Z = \Tr \left[ e^{-\beta H} \right] = \Tr_{\text{spin}} \Tr_{\text{phot}} \left[ D^{\dagger} e^{-\beta (H_{\text{eff}} + \omega a^{\dagger}a)} D \right].
\end{equation}
Using the cyclic property of the trace ($\Tr[XYZ] = \Tr[ZXY]$) within the photon subspace, the unitary operators cancel each other ($D D^{\dagger} = 1$):
\begin{equation}
    Z = \Tr_{\text{spin}} \left[ e^{-\beta H_{\text{eff}}} \Tr_{\text{phot}} \left[ D D^{\dagger} e^{-\beta \omega a^{\dagger}a} \right] \right], 
\end{equation}
where we used $[a^{\dagger}a, H_{\text{eff}}] = 0$.
Since $H_{\text{eff}}$ acts only on the spin Hilbert space and $\omega a^\dagger a$ acts only on the photon Hilbert space, the trace fully factorizes:
\begin{equation}
    Z = \underbrace{\Tr_{\text{spin}} \left[ e^{-\beta H_{\text{eff}}} \right]}_{Z_{\text{spin}}} \times \underbrace{\Tr_{\text{phot}} \left[ e^{-\beta \omega a^{\dagger}a} \right]}_{Z_{\text{boson}}}.
\end{equation}
This result proves that the photon part $Z_{\text{boson}} = (1 - e^{-\beta\omega})^{-1}$ is a purely multiplicative factor independent of the spin configuration. Therefore, for all thermodynamic calculations, the operator $b$ acts as a decoupled free boson, and the system is governed solely by the effective spin Hamiltonian $H_{\text{eff}}$.

\section{General Photon Statistics}
\label{sec:photon_stats}

In this section, we derive the general expressions for the photon observables. As shown in Sec.~\ref{sec:general_theory}, the unitary displacement transformation decouples the bosonic mode $b$ from the spin degrees of freedom. The original photon operator $a$ is related to the transformed boson operator $b$ and the collective spin operator $S^\mu \equiv \sum_{i=1}^N \sigma_i^\mu$ by the relation:
\begin{equation}
    a = b - \frac{g}{\omega\sqrt{N}} S^\mu.
    \label{eq:a_b_relation}
\end{equation}
Since the transformed Hamiltonian $H_{\text{eff}}$ contains no interaction between $b$ and the spins, the thermal expectation value of $b$ vanishes ($\langle b \rangle = 0$).

\subsection{Cavity Field Order Parameter $\langle a \rangle$}
The expectation value of the cavity field amplitude $\langle a \rangle$ serves as the primary order parameter. Taking the thermal average of Eq.~(\ref{eq:a_b_relation}), we obtain:
\begin{equation}
    \langle a \rangle = \langle b \rangle - \frac{g}{\omega\sqrt{N}} \langle S^\mu \rangle = -\frac{g}{\omega\sqrt{N}} \langle S^\mu \rangle.
\end{equation}
Defining the order parameter $m$ associated with the collective spin as $m \equiv \frac{1}{N}\langle S^\mu \rangle$, we arrive at the exact relation:
\begin{equation}
    \frac{\langle a \rangle}{\sqrt{N}} = -\frac{gm}{\omega}.
\end{equation}

\subsection{Normalized Photon Number $n$}
The normalized photon number $n \equiv \langle a^\dagger a \rangle / N$ is another key observable. Using Eq.~(\ref{eq:a_b_relation}), we calculate the operator product $a^\dagger a$:
\begin{equation}
    a^\dagger a = \left( b^\dagger - \frac{g}{\omega\sqrt{N}} S^\mu \right) \left( b - \frac{g}{\omega\sqrt{N}} S^\mu \right).
\end{equation}
Expanding this product and using $\langle b \rangle = \langle b^\dagger \rangle = 0$, the cross-terms vanish. We obtain:
\begin{equation}
    \langle a^\dagger a \rangle = \langle b^\dagger b \rangle + \frac{g^2}{N\omega^2} \langle (S^\mu)^2 \rangle.
    \label{eq:adaggera_exact}
\end{equation}
The first term $\langle b^\dagger b \rangle$ follows the Bose-Einstein distribution $(e^{\beta\omega}-1)^{-1}$. 
The second term involves the square of the collective spin operator, which can be expanded as:
\begin{equation}
    (S^\mu)^2 = \left( \sum_{i=1}^N \sigma_i^\mu \right)^2 = \sum_{i=1}^N (\sigma_i^\mu)^2 + \sum_{i \ne j} \sigma_i^\mu \sigma_j^\mu.
\end{equation}
Using $(\sigma_i^\mu)^2 = 1$, the first term is simply $N$. For the second term (the all-to-all interaction part), we apply the \textbf{partial mean-field approximation} introduced in the main text. This approximation replaces the spin-spin correlation with the product of the mean fields:
\begin{equation}
    \sum_{i \ne j} \sigma_i^\mu \sigma_j^\mu \simeq \sum_{i \ne j} m^2 \simeq N^2 m^2.
\end{equation}
Substituting this back into Eq.~(\ref{eq:adaggera_exact}), we obtain:
\begin{equation}
    \langle a^\dagger a \rangle \simeq \frac{1}{e^{\beta\omega}-1} + \frac{g^2}{N\omega^2} (N + N^2 m^2).
\end{equation}
Finally, dividing by $N$ to define the normalized photon number $n$, and taking the thermodynamic limit ($N \to \infty$) where terms of order $1/N$ vanish, we arrive at:
\begin{equation}
    n \simeq \frac{g^2 m^2}{\omega^2}.
\end{equation}
This derivation explicitly demonstrates that the photon number relation relies on the same partial mean-field approximation used to solve the spin Hamiltonian.

\section{Exact Solution for the 1D Ising Model}
\label{sec:ising_exact}

We now apply the general theory to the 1D ferromagnetic Ising chain ($H_{\text{spin}} = -J \sum_{i} \sigma_i^z \sigma_{i+1}^z$) coupled via the $z$-component ($\mu=z$). The effective Hamiltonian becomes:
\begin{equation}
    H_{\text{eff}} = -J\sum_{i=1}^{N}\sigma_{i}^z\sigma_{i+1}^z - \frac{g^{2}}{N\omega}\sum_{i \ne j}\sigma_{i}^z\sigma_{j}^z .
    \label{eq:heff_ising}
\end{equation}
The second term is an all-to-all interaction. We prove that the mean-field treatment of this term (partial mean-field approximation) is exact in the thermodynamic limit.

\subsection{Hubbard-Stratonovich Transformation}
The partition function of the spin part is $Z_{\text{spin}} = \Tr_{\sigma} \exp\left[ \beta J \sum_i \sigma_i^z \sigma_{i+1}^z + \frac{\beta g^2}{N\omega} (\sum_i \sigma_i^z)^2 \right]$. We employ the Hubbard-Stratonovich (HS) transformation~\cite{hubbard1959calculation,stratonovich1957method}:
\begin{equation}
    e^{\alpha A^2} = \sqrt{\frac{\alpha}{\pi}} \int_{-\infty}^{\infty} dx \, e^{-\alpha x^2 + 2\alpha x A}.
\end{equation}
Setting $\alpha = N\beta g^2/\omega$ and $A = (1/N)\sum_i \sigma_i^z$, and rescaling $x=m$, the partition function becomes an integral over an auxiliary field $m$:
\begin{equation}
    Z_{\text{spin}} \propto \int_{-\infty}^{\infty} dm \, e^{-N\beta \frac{g^2}{\omega} m^2} \Tr_{\sigma} \exp\left[ \beta J \sum_i \sigma_i^z \sigma_{i+1}^z + \beta \left( \frac{2g^2 m}{\omega} \right) \sum_i \sigma_i^z \right].
\end{equation}
The term inside the trace is the partition function of a 1D Ising chain in an effective field $h_{\text{eff}} = 2g^2 m / \omega$, denoted as $Z_{\text{1D}}(h_{\text{eff}})$. Thus,
\begin{equation}
    Z_{\text{spin}} \propto \int_{-\infty}^{\infty} dm \, \exp\left[ -N\beta \left( \frac{g^2}{\omega} m^2 - \frac{1}{N\beta} \ln Z_{\text{1D}}(h_{\text{eff}}) \right) \right].
\end{equation}

\subsection{Saddle-Point Approximation and Exactness}
In the thermodynamic limit ($N \to \infty$), the integral is dominated by the saddle point of the effective free energy density $F(m) = \frac{g^2}{\omega} m^2 + f_{\text{1D}}(h_{\text{eff}})$, where $f_{\text{1D}}(h) = -(N\beta)^{-1} \ln Z_{\text{1D}}(h)$. The saddle-point condition $\partial F / \partial m = 0$ yields:
\begin{equation}
    \frac{2g^2 m}{\omega} + \frac{\partial f_{\text{1D}}(h_{\text{eff}})}{\partial h_{\text{eff}}} \frac{\partial h_{\text{eff}}}{\partial m} = 0.
\end{equation}
Using $\partial h_{\text{eff}} / \partial m = 2g^2/\omega$ and $m_{\text{1D}}(h) = -\partial f_{\text{1D}}(h) / \partial h$, we obtain the self-consistent equation:
\begin{equation}
    m = m_{\text{1D}}\left( h = \frac{2g^2 m}{\omega} \right).
\end{equation}
This confirms that the order parameter $m$ is determined by the magnetization of the reference 1D system in the self-consistent field.

\subsection{Explicit Solution}
Using the exact solution for the 1D Ising model~\cite{ising1925}, the magnetization is:
\begin{equation}
    m_{\text{1D}}(h) = \frac{\sinh(\beta h)}{\sqrt{\sinh^2(\beta h) + e^{-4\beta J}}}.
\end{equation}
Substituting $h = 2g^2 m / \omega$, we arrive at the exact self-consistent equation:
\begin{equation}
    m = \frac{\sinh\left( 2\beta \frac{g^2}{\omega} m \right)}{\sqrt{\sinh^2\left( 2\beta \frac{g^2}{\omega} m \right) + e^{-4\beta J}}}.
    \label{eq:sm_final_eq}
\end{equation}

\section{Derivation of the Critical Temperature $T_c$}
\label{sec:critical_temp}

The critical temperature $T_c$ is defined as the temperature at which the trivial solution $m=0$ becomes unstable. We expand the right-hand side of Eq.~(\ref{eq:sm_final_eq}) for small $m$. Using $\sinh(x) \simeq x$ and $e^y \simeq 1$, we obtain:
\begin{equation}
    m \simeq \frac{2\beta \frac{g^2}{\omega} m}{\sqrt{e^{-4\beta J}}} = m \left( \frac{2g^2}{\omega T} e^{2J/T} \right).
\end{equation}
A non-trivial solution emerges when the coefficient equals unity:
\begin{equation}
    1 = \frac{2g^2}{\omega T_c} e^{2J/T_c}.
    \label{eq:Tc_implicit}
\end{equation}
Let $X = 2J/T_c$. Substituting $T_c = 2J/X$ into Eq.~(\ref{eq:Tc_implicit}):
\begin{equation}
    X e^X = \frac{\omega J}{g^2}.
\end{equation}
Using the Lambert $W$ function ($W(z) e^{W(z)} = z$), the solution is $X = W(\omega J / g^2)$.
Thus, the explicit formula for $T_c$ is:
\begin{equation}
    T_c = \frac{2J}{W\left( \frac{\omega J}{g^2} \right)}.
\end{equation}

\end{document}